\documentclass[twocolumn]{aastex63}

\usepackage[super]{nth}
\usepackage{my_astro}
\usepackage{needspace}
\usepackage{aas_macros}

\newcommand{\lrz}{\log_{10}R_0}
\newcommand{\lr}{\log_{10}R}
\newcommand{\lrgap}{\log_{10}R_\mathrm{gap}}

\newcommand{\lratioN}{$L_{NUV_e}/L_\mathrm{bol}$}

\newcommand{\lratioX}{$L_\mathrm{XUV}/L_\mathrm{bol}$}
\newcommand{\lratioXm}{L_\mathrm{XUV}/L_\mathrm{bol}}
\newcommand{\xNUV}{$NUV_e$}
\newcommand{\xFUV}{$FUV_e$}
\newcommand{\IxNUV}{$I_{NUV_e}$}
\newcommand{\IxNUVm}{I_{NUV_e}}
\newcommand{\IxFUV}{$I_{FUV_e}$}
\newcommand{\IXUV}{$I_{XUV}$}

\newcommand{\teff}{T_\mathrm{eff}}
\newcommand{\M}{$M_\star$}
\newcommand{\Mm}{M_\star}
\newcommand{\R}{R_p}

\newcommand{\fitCaptionBoiler}{The blue line and translucent area are the best fit to the gap and bootstrapped $1\sigma$ uncertainty. In the right panel, the best fit has been subtracted in log space, aligning the gap to the histogram.}

\graphicspath{{./}{figures/}}

\received{2019 Oct 22}
\revised{2019 Dec 14}
\accepted{2019 Dec 23}
\submitjournal{The Astrophysical Journal}

\shorttitle{Stellar Activity, Stellar Mass, and the Exoplanet Radius Gap}
\shortauthors{Loyd et al.}

\begin{document}

\title{Current Population Statistics Do Not Favor Photoevaporation over Core-Powered Mass Loss as the Dominant Cause of the Exoplanet Radius Gap}

\correspondingauthor{R. O. Parke Loyd}
\email{parke@asu.edu}

\author[0000-0001-5646-6668]{R. O. Parke {Loyd}}
\affiliation{School of Earth and Space Exploration, Arizona State University, Tempe, AZ 85287}

 \author[0000-0002-7260-5821]{Evgenya L. {Shkolnik}}
 \affiliation{School of Earth and Space Exploration, Arizona State University, Tempe, AZ 85287}

 \author[0000-0002-6294-5937]{Adam C. {Schneider}}
 \affiliation{School of Earth and Space Exploration, Arizona State University, Tempe, AZ 85287}

 \author[0000-0003-1290-3621]{Tyler {Richey-Yowell}}
 \affiliation{School of Earth and Space Exploration, Arizona State University, Tempe, AZ 85287}

 \author[0000-0002-7129-3002]{Travis S. {Barman}}
 \affiliation{Lunar and Planetary Laboratory, University of Arizona, Tucson, AZ 85721 USA}

 \author[0000-0002-1046-025X]{Sarah {Peacock}}
 \affiliation{Lunar and Planetary Laboratory, University of Arizona, Tucson, AZ 85721 USA}

 \author[0000-0001-9573-4928]{Isabella {Pagano}}
 \affiliation{INAF - Osservatorio Astrofisico di Catania, Via S. Sofia 78, 95123, Catania, Italy}


\begin{abstract}
We search for evidence of the cause of the exoplanet radius gap, i.e. the dearth of planets with radii near $1.8$~\Rearth. 
If the cause was photoevaporation, the radius gap should trend with proxies for the early-life high-energy emission of planet-hosting stars.
If, alternatively, the cause was core-powered mass loss, no such trends should exist.
Critically, spurious trends between the radius gap and stellar properties arise from an underlying correlation with instellation.
After accounting for this underlying correlation, we find no trends remain between the radius gap and stellar mass or present-day stellar activity as measured by near-UV emission. 
We dismiss the nondetection of a radius gap trend with near-UV emission because present-day near-UV emission is unlikely to trace early-life high-energy emission, but we provide a catalog of \textit{GALEX} near-UV and far-UV emission measurements for general use.
We interpret the nondetection of a radius gap trend with stellar mass by simulating photoevaporation with mass-dependent evolution of stellar high-energy emission. 
The simulation produces an undetectable trend between the radius gap and stellar mass under realistic sources of error.
We conclude that no evidence, from this analysis or others in the literature, currently exists that clearly favors either photoevaporation or core powered mass loss as the primary cause of the exoplanet radius gap. 
However, repeating this analysis once the body of well-characterized $< 4\ R_\oplus$ planets has roughly doubled could confirm or rule out photoevaporation.
\end{abstract}


\needspace{6em} \section{Introduction}
\label{sec:intro}

The population of $< 4\ R_\oplus$ planets is split into two distinct groups.
The two groups are separated by a sharp drop in the occurrence rate of planets with radii near $1.8\ R_\oplus$ \citep{fulton17}, often termed the radius gap.
As the precision of planetary radius measurements has improved, now reaching 10\% or better for around 1700 $< 4\ R_\oplus$ planets, the radius gap has become increasingly clear \citep{owen13,fulton17,fulton18,eylen18,martinez19,macdonald19}.

The radius gap probably represents a sharp division between planets that retain a thick, primordial H/He atmosphere and planets that either lost their thick atmosphere or formed without one \citep{owen13,lopez13,ginzburg16,fulton18}.
A planet with an H/He atmosphere comprising roughly 1\% of its mass will have an apparent radius in visible light several times that of its rocky core \citep{lopez14}.
Below a critical core mass, a planet's gravity will  be too weak to retain its primordial atmosphere.
The apparent radius of a planet that cannot retain its primordial atmosphere shrinks as atmospheric mass is lost.
The difference in the apparent radius of cores that have retained thick atmospheres and cores that have lost them gives rise to the radius gap.
Planets above the radius gap have lower bulk densities than those below the radius gap, supporting the idea that those above have thick atmospheres and those below do not \citep{weiss14,swain18}.

Several mechanisms could strip primordial atmospheres from rocky planetary cores.
Under the XUV evaporation theory, stellar extreme ultraviolet and X-ray (XUV, $<912$~\AA) emission drives most of the atmospheric mass loss within the first 100~Myr \citep{owen12,lopez12}.
Alternatively, under the core-powered mass loss theory, the residual accretion heat of the planetary core powers atmospheric loss over the course of a Gyr or more \citep{ginzburg16,ginzburg18}.
Meanwhile, exoplanet demographics do not support theories invoking differences in protoplanetary disk dispersal and giant impacts \citep{fulton18}.
The XUV evaporation and core-powered mass loss theories have a key difference: XUV evaporation relies on stellar activity, whereas core-powered mass loss is indifferent to it.
In this work, we leveraged that difference to better understand what causes the radius gap.

\subsection{Details of the XUV Evaporation and Core-Powered Mass Loss Theories}
Under the XUV evaporation theory, atmospheric loss depends fundamentally on the XUV irradiation of the planet by the host star, $I_\mathrm{XUV}$, in its early life \citep{owen12,lopez12}.
As $I_\mathrm{XUV}$ increases, more massive planetary cores can be stripped of their primordial atmospheres.
XUV irradiation is the product of two components, the planet's instellation, $S$, and the fraction of the star's bolometric luminosity emitted at XUV wavelengths, \lratioX,  \begin{equation}
    \label{eqn:S}
    I_\mathrm{XUV} = S\cdot L_\mathrm{XUV}/L_\mathrm{bol}.
\end{equation}
Instellation is a generalization of the term insolation used for solar system planets. 
It indicates the bolometric stellar flux at the position of the planet, i.e., $L_\mathrm{bol}/4\pi a^2$ where $L_\mathrm{bol}$ is the stellar bolometric luminosity and a is the semi-major axis of the planet's orbit.
Because $I_\mathrm{XUV} \propto S$, XUV evaporation predicts an increase in the radius gap with $S$.
Throughout this paper, when discussing increases, decreases, trends, etc. of the radius gap, we refer to changes in the location of the gap on the planetary radius axis, i.e., the local minimum in a histogram of planetary radii.
Because shorter period planets are more irradiated, $S \propto P^{-4/3}$, XUV evaporation further predicts a decrease in the radius gap with increasing orbital period.
These trends have been confirmed in multiple analyses of planetary demographics \citep{fulton17,fulton18,eylen18,martinez19,macdonald19}.

Alternatively, in the core-powered mass loss framework, atmospheric loss depends fundamentally on the planetary equilibrium temperature, $T_\mathrm{eq}$.
In this framework, the opaque, thick primordial atmosphere of a newly-formed rocky planet prevents the core from quickly radiating its leftover heat of formation to space.
Instead of being radiated away, this heat powers a Parker wind \citep{ginzburg16,ginzburg18}. 
The sonic point of the wind throttles the outflow, limiting it to the sound speed.
The sound speed at the sonic point increases with planetary equilibrium temperature, enabling greater rates of atmospheric escape.
Hence, the rate of atmospheric loss is controlled by the host star, even though the star does not directly power it.
Because $T_\mathrm{eq} \propto S^{1/4}$, core-powered mass loss predicts the same qualitative dependencies on $S$ and $P$ as XUV evaporation.

The similarity of the predictions of core-powered mass loss and XUV evaporation makes finding lines of evidence that could disentangle the two challenging.
Both can reproduce the observed planetary population in the radius, instellation, and orbital period parameter space  \citep{owen17,gupta19a,gupta19b,wu19}.
A key to disentangling the two could be \lratioX.
Core-powered mass loss does not depend on \lratioX, whereas XUV evaporation depends fundamentally on it.
Therefore, a trend of the observed radius gap with \lratioX\ would be evidence for XUV evaporation.

\subsection{An Overview of Stellar XUV Evolution}
\label{sec:evol}

Initially, a star's XUV emission is saturated at a constant level.
Ratios of X-ray to bolometric emission during saturation are $\sim10^{-3}$--$10^{-4}$, increasing toward later type stars \citep{jackson12}.
Saturation lasts anywhere from tens of Myr to several Gyr, also increasing toward later stellar types (e.g., \citealt{west08}).
Afterwards, XUV emission declines roughly as $t^{-1}$ (e.g. \citealt{jackson12,shkolnik14}).
The stellar mass dependency of both early life and lifetime-integrated XUV emission roughly goes as $\lratioXm \propto \Mm^{-2}$ in the recent empirical model of \cite{mcdonald19}.
Because XUV emission is highest when stars are young, under XUV evaporation most of the atmospheric stripping a planet experiences occurs within its first 100~Myr \citep{owen17}.
Compounding this effect is the hot and distended state of planetary atmospheres in their youth \citep{luger15}.

Although XUV emission evolves smoothly in population averages, differences between individual stars are large.
Scatter in seed rotation rates of otherwise identical stars could spread their saturation lifetimes over an order of magnitude \citep{tu15}.
Meanwhile, observed emission from stars of similar age, mass, and rotation rates at X-ray, FUV, and NUV wavelengths exhibit 1$\sigma$ scatter of 0.4-0.5~dex, well beyond measurement uncertainties \citep{jackson12,shkolnik14,schneider18,france18,richey19,macdonald19}.
This is true even during the saturation phase, when population-averaged activity values remain constant.
Differences in XUV emission between stars might measurably influence the evaporation of their planets \citep{kubyshkina19}.

\subsection{Mass and Present NUV Activity as Proxies for a Star's Early-life XUV Emission}
A star's \lratioX\ cannot be directly measured.
The portion of the XUV spanning $\sim$400-912~\AA\ is subject to strong absorption by \ion{H}{1} in the interstellar medium.
Supposing \lratioX\ could be measured, the measurement would only be a snapshot of the star's present-day emission, not the early-life emission that is critical to XUV evaporation.
Under these limitations, the only recourse is to seek proxies for early-life \lratioX.

One proxy for early-life \lratioX\ of stars is stellar mass (Section \ref{sec:evol}).
\cite{zeng17}, \cite{fulton18}, \cite{wu19}, and \cite{gupta19b} all examined the radius gap as a function of \M.
Our analysis is novel in that we use tests that account for differences in $S$ when searching for a dependency on \M.
The importance of this accounting, and the results from previous work, will be discussed in the context of the new results in Section \ref{sec:discuss}.

A second possible proxy for early-life \lratioX\ is present-day stellar activity.
An archive of readily-accessible activity measurements exists in the form of the all-sky NUV and FUV observations made by the \textit{Galaxy Evolution Explorer} (\textit{GALEX}) spacecraft.
However, using \textit{GALEX} archival data, or any present-day measurements of stellar activity, comes with a critical caveat.
For \textit{GALEX} NUV or FUV emission to serve as a useful axis on which to examine the radius gap, differences in the present day NUV or FUV emission of individual stars must trace back to differences in their early-life XUV emission.
A dispersion of $\sim$0.5~dex in X-ray activity for young stars remains just as large for field-age stars \citep{jackson12}, even though, over the course of their evolution, the rotation periods of solar-mass stars converge to very similar values \citep{epstein14}.
This hints that variations in activity between individuals might not be tied purely to differences in rotation and could persist throughout the rotational evolution of stellar populations. 
If so, there could be a fossilized dependency of the radius gap on present-day stellar activity.

We structure this paper as follows.
In Section \ref{sec:data}, we give a description of the exoplanet catalog we used, a catalog of \textit{GALEX} NUV and FUV activity measurements we assembled, and our method for fitting the radius gap.
Section \ref{sec:results} presents several tests for relationships between the radius gap, stellar activity, and stellar mass and the results of those tests.
A discussion of those results follows in Section \ref{sec:discuss}, and we conclude with a brief summary in Section \ref{sec:summary}.
Readers interested in the details of the data and various analyses, particularly those wishing to reproduce any portion of this work, will find those details in the Appendices.

\needspace{6em} \section{Data and Analysis\label{sec:data}}
The population of planets we analyzed consists of the confirmed and \textit{Kepler} candidate exoplanet systems downloaded from the NASA Exoplanet Archive on 2019 June 7.
We updated system parameters to those of the \textit{California Kepler Survey} \citep{fulton18} and the asteroseismic survey of \cite{eylen18} where possible.
These surveys include precise estimates of stellar masses and planetary radii that clearly expose the radius gap.
We obtained archival NUV and FUV photometry of all systems observed by the \textit{GALEX} spacecraft and estimated the fraction of that emission attributable to stellar activity (details in Section \ref{sec:uv}).
A portion of the catalog of \textit{GALEX} fluxes is presented in Table \ref{tbl:galex}, and the full table will be available with the published version of this article. 

\begin{figure}
\includegraphics{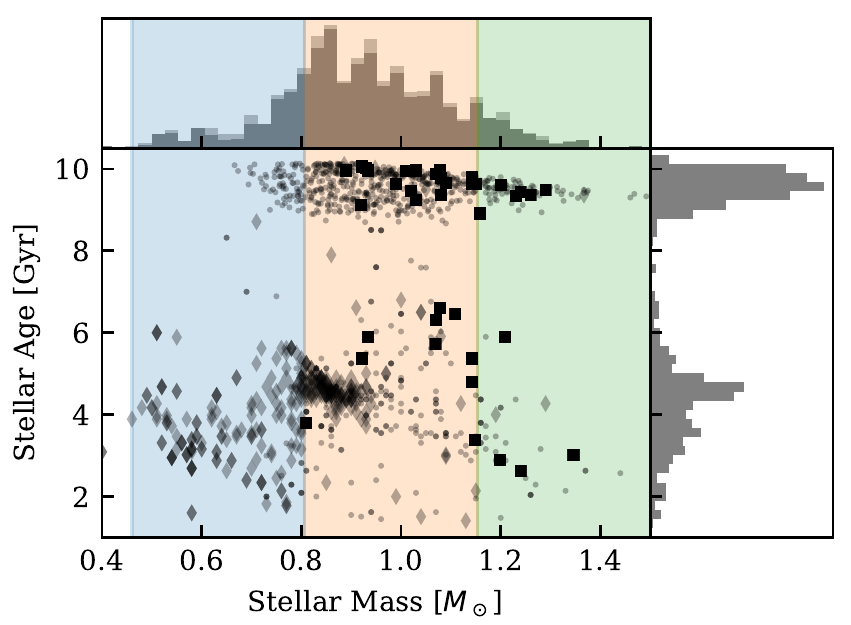}
\caption{Distribution of ages and host star masses for the 1234 systems with cataloged values for stellar mass, planetary instellation, and system age.
The light gray extensions to the top histogram include systems that have a catalog value for stellar mass and planetary instellation, but not system age; 1365 systems total.
For multi-planet systems, the host star is counted once for each planet.
The parameters of circular points come from \cite{fulton18}, squares from \cite{eylen18}, and diamonds from the NASA Exoplanet Archive.
The three colored regions correspond to stellar mass bins referenced later in this paper (Section \ref{sec:mass}).
\label{fig:mass_age_corr}}
\end{figure}

We applied the following cuts to the exoplanet sample:
\begin{itemize}
    \item Stellar $\log_{10} g$ in the range (4, 5)~$\mathrm{cm\ s^{-2}}$ to exclude evolved stars,
    \item Planet $R_p < 4\ R_\oplus$,
    \item Planet impact parameter $b < 0.9$ to avoid strong correlations between planetary radius and impact parameter uncertainties \citep{fulton18},
    \item Planet orbital period $P < 100$ d to exclude planets where stellar irradiation might not be sufficient to create a radius gap (e.g., \citealt{owen17}),
    \item Precision on the planet radius of $\sigma_{R_p}/R_p < 0.1$ to mitigate scatter of points into the radius gap resulting from random measurement error.
\end{itemize}
This yielded a sample of 1548 planets (in 1105 unique systems), 942 from the \textit{California Kepler Survey} \citep{fulton18} and 1365 with measurements of stellar mass and planetary instellation (in 958 unique systems).
Figure \ref{fig:mass_age_corr} depicts the distribution of catalog masses and ages of the sample's host stars.
Age estimates in Figure \ref{fig:mass_age_corr} are mostly isochrone estimates with uncertainties of order Gyr.
We were able to estimate the activity-related NUV irradiation of 697 planets and the activity-related FUV irradiation of 176 planets, with upper limits for the NUV irradiation of 189 planets and FUV irradiation of 544 planets.

Our investigation required quantifying the relationship between the radius gap and several independent variables, such as instellation.
We assumed a linear relationship between the radius gap and independent variable in log-log space, in line with previous works (e.g. \citealt{eylen18,martinez19}).
To fit the linear gap, we created an algorithm that subtracted the radius gap from the planetary radii, then used kernel density estimation (KDE; akin to a smoothed histogram) to estimate the number of planets along the line of the gap.
We took the gap line with the minimum number density of planets along it as the best fit.
By bootstrapping the planet sample, we determined the median fit parameters and their uncertainties (more details in Appendix \ref{app:fits}).

We did not attempt to correct planet occurrence rates according to detection completeness, in line with \cite{eylen18}, among others.
The location of the radius gap does not appreciably change between studies that include and exclude completeness corrections \citep{fulton18,eylen18}.

\subsection{\textit{GALEX} NUV and FUV Fluxes of Known Exoplanet Host Stars}
\label{sec:uv}

\textit{GALEX} observed nearly the entire sky during its operation from 2003-2012.
The observatory conducted photometry in two broad UV bands, an FUV band covering roughly 1350-1800~\AA\ and an NUV band covering roughly 1700-3000~\AA.
We obtained NUV and FUV fluxes of the host stars, or upper limits, from the \textit{GALEX} master source catalog (MCAT) hosted by the Mikulski Archive for Space Telescopes (MAST).
Although only planets with radii measurements (i.e., transiting planets) were relevant to this paper, we compiled values for all confirmed and candidate systems for the benefit of the community.
Observations of most objects included two to ten exposures of a few hundred seconds each and typically spanned several years.
Thus, the measurements for most stars sampled a variety of stellar rotation states and a fraction of a Sun-like activity cycle.

We created a simple \textit{Python} tool to pull \textit{GALEX} fluxes from the visit level archive hosted by MAST while correcting for target proper motions.\footnote{\url{https://github.com/parkus/galex_motion}}
We imposed a maximum 16'' distance to match targets. 
For targets with multiple exposures, we averaged the exposure-weighted count rates after removing any outliers.
When no source existed within the match radius in the \textit{GALEX} source catalog, we computed 2$\sigma$ upper limits based on the background count rate reported in the catalog.
When \textit{GALEX} covered the target's coordinates with multiple visits, we used the most restrictive upper limit yielded by any single visit.
Our catalog adds 2987 new planet host stars with \textit{GALEX} photometry to the 272 compiled by \cite{shkolnik13}.
We vetted our fluxes against those of \cite{shkolnik13} and investigated anomalies as described in Appendix \ref{app:vetting}.
\textit{GALEX} measurements of bright targets are subject to saturation effects, and we corrected for these effects using a polynomial fit to the detector response published in \cite{morrissey07}.

\begin{figure}
\includegraphics{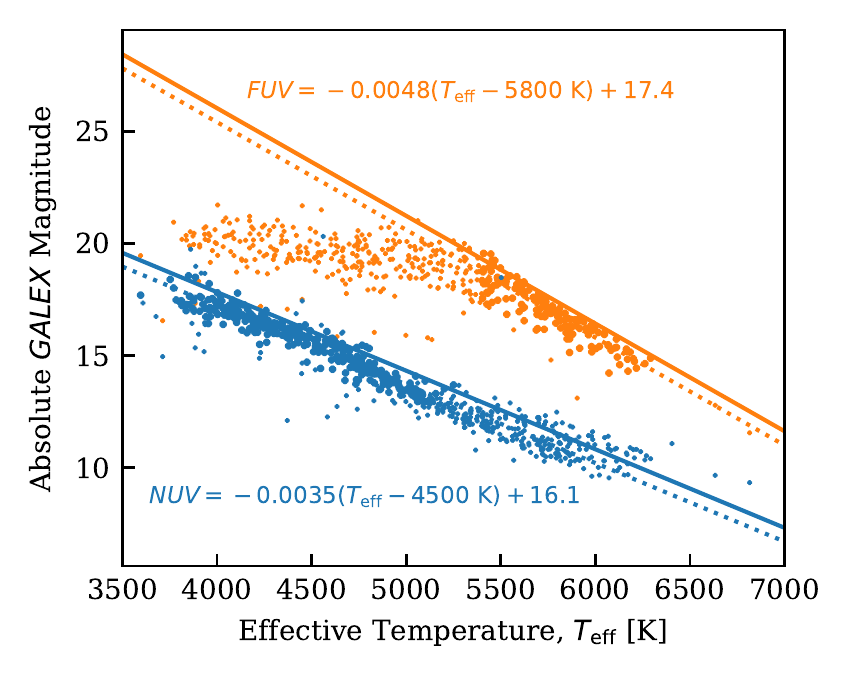}
\caption{Fits to the \textit{GALEX} magnitudes of the CATSUP sample.
We excluded smaller points from the fits for a variety of reasons (see text).
Dotted lines show the fits, whereas solid lines give the adopted photosphere-only limit subtracted to give a star's excess \textit{GALEX} flux.
The knee in the FUV points results from stellar activity contributing more emission than the photosphere for low-mass stars.
Equations are for the solid lines.
\label{fig:catsup}}
\end{figure}

To make the measurements of host-star NUV and FUV emission the best possible proxies for activity-generated XUV emission, we estimated the fraction of NUV and FUV emission attributable to stellar activity.
We did this by subtracting an estimate of each star's photospheric emission in the FUV and NUV bands.
To estimate photospheric emission, we relied on the CATSUP sample (Figure \ref{fig:catsup}; \citealt{hinkel17}).
For the stars in that sample, we assumed those with the lowest FUV and NUV emission represent a limit where photospheric emission dominates.
We fit a power law between effective temperature and absolute FUV and NUV magnitudes, then found the offset that placed the power law at the \nth{95} percentile of the data points.
This we took as an estimate of the contribution of the photosphere to FUV and NUV emission for a star of a given effective temperature.
The FUV fluxes of stars with $\teff < 5300$~K were all dominated by activity and the NUV emission of stars with $\teff > 5200$~K generally exceeded the linearity limit for the \textit{GALEX} detector.
We excluded these points from the fit, along with any points that were upper or lower limits and $>$2.5$\sigma$ outliers.
By subtracting the photospheric fit from the \textit{GALEX} fluxes we retrieved for each exoplanet host star, we obtained the excess FUV and NUV flux attributable to activity, \xFUV\ and \xNUV.
We denote the irradiation of orbiting planets by excess FUV and NUV flux as \IxFUV\ and \IxNUV.

\begin{deluxetable*}{lrrrrrrrrrrr}
\tablewidth{0pt}
\tabletypesize{\footnotesize}
\rotate
\tablecaption{Host star \textit{GALEX} fluxes and planetary irradiations for all known planetary systems.\label{tbl:galex}}
\tablehead{\colhead{Planet} & \colhead{\textit{NUV}} & \colhead{\textit{NUV}} & \colhead{Excess \textit{NUV}\tablenotemark{a}} & \colhead{\textit{FUV}} & \colhead{\textit{FUV}} & \colhead{Excess \textit{FUV}\tablenotemark{a}} & \colhead{Orbital} & \colhead{Bolometric} & \colhead{Stellar} & \colhead{Stellar}\\
\colhead{} & \colhead{} & \colhead{at planet} & \colhead{at planet} & \colhead{} & \colhead{at planet} & \colhead{at planet} & \colhead{Period} & \colhead{Irradiation} & \colhead{Mass} & \colhead{Luminosity}\\
\colhead{} & \colhead{$\mathrm{erg\ cm^{-2}\ s^{-1}}$} & \colhead{$\mathrm{erg\ cm^{-2}\ s^{-1}}$} & \colhead{$\mathrm{erg\ cm^{-2}\ s^{-1}}$} & \colhead{$\mathrm{erg\ cm^{-2}\ s^{-1}}$} & \colhead{$\mathrm{erg\ cm^{-2}\ s^{-1}}$} & \colhead{$\mathrm{erg\ cm^{-2}\ s^{-1}}$} & \colhead{d} & \colhead{$S_\oplus$} & \colhead{$M_\odot$} & \colhead{$L_\odot$}}
\startdata
11 Com b & 3.5$\times10^{-11}$ & 1700 & 1600 & 2.0$\times10^{-13}$ & 44 & 44 & 326.03 & 105.17 & 2.70 & 174.98\\
HD 24064 b & 1.8$\times10^{-12}$ & 3900 & 3900 & 1.9$\times10^{-14}$ & 44 & 44 & 535.60 & 211.29 & 1.61 & 351.56\\
KOI-1599.01 & \nodata & \nodata & \nodata & \nodata & \nodata & \nodata & 20.44 & 55.69 & 1.02 & 1.21\\
Kepler-1468 c & \nodata & \nodata & \nodata & \nodata & \nodata & \nodata & 3.55 & 1479.57 & 1.04 & 3.22\\
Kepler-300 b & 1.5$\times10^{-13}$ & 9.0$\times10^{5}$ & 5.7$\times10^{5}$ & $<$4.8$\times10^{-14}$ & 7.7$\times10^{5}$ & 7.7$\times10^{5}$ & 10.45 & 265.19 & 1.07 & 2.05\\
Kepler-672 b & 2.1$\times10^{-13}$ & \nodata & \nodata & 3.7$\times10^{-14}$ & \nodata & \nodata & 38.38 & 16.36 & 0.95 & 0.76\\
WASP-23 b & 1.8$\times10^{-13}$ & 2.4$\times10^{5}$ & 98000 & \nodata & \nodata & \nodata & 2.94 & \nodata & 0.78 & 0.42\\
K02807.01 & 3.9$\times10^{-13}$ & 9.3$\times10^{6}$ & 3.9$\times10^{6}$ & \nodata & \nodata & \nodata & 3.38 & 1111.66 & 1.31 & \nodata\\
K04750.01 & \nodata & \nodata & \nodata & \nodata & \nodata & \nodata & 36.72 & 14.51 & 0.94 & \nodata\\
K05960.01 & 1.1$\times10^{-12}$ & $<$24000 & $<$-1600 & \nodata & \nodata & \nodata & 32.42 & 52.70 & 0.84 & \nodata\\
\enddata

\tablenotetext{a}{Flux remaining after the subtraction of an empirical estimate of the flux from the star's photosphere.}

\tablecomments{Only 10 planets from the full catalog are shown here, with uncertainties and references excluded due to space constraints. The full catalog will be available as a machine-readable table in the published version of the article.}
\end{deluxetable*}

\needspace{6em} \section{Results}
\label{sec:results}
Our fits to the radius gap are consistent with those reported in previous works \citep{fulton18,eylen18,martinez19}.
We find the slope of the gap in the $R_p$-$P$ plane to be $-0.08\pm0.01$, in comparison to $-0.09^{+0.02}_{-0.04}$ \citep{eylen18} and $-0.11\pm0.02$ \citep{martinez19}.
Fitting subsets of the population causes the slope of the gap in $R_p$-$P$ to vary between -0.05 and -0.10, though -0.08 remains within 1$\sigma$ of each fit.

\begin{figure*}
\includegraphics{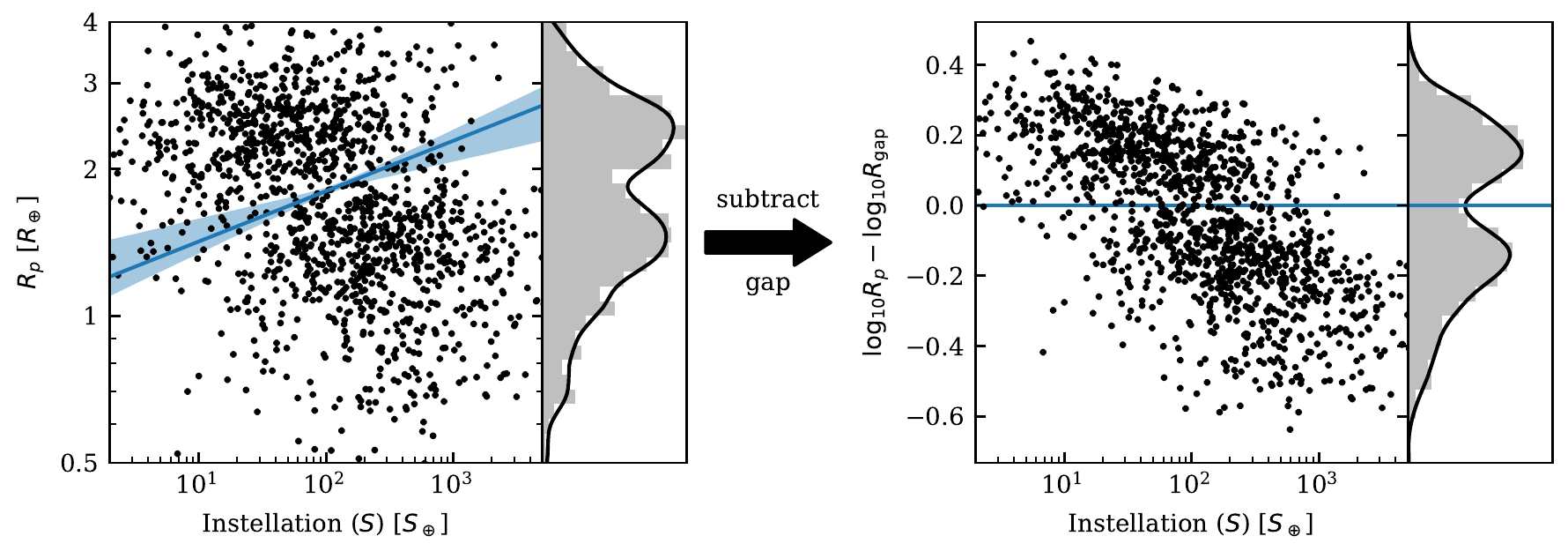}
\caption{Planetary radius gap in the $R_p$-S (instellation) plane.
\fitCaptionBoiler\
There is a positive relationship between the radius gap and instellation, consistent with previous results (e.g., \citealt{fulton18}).
\label{fig:S}}
\end{figure*}

\begin{figure*}
\includegraphics{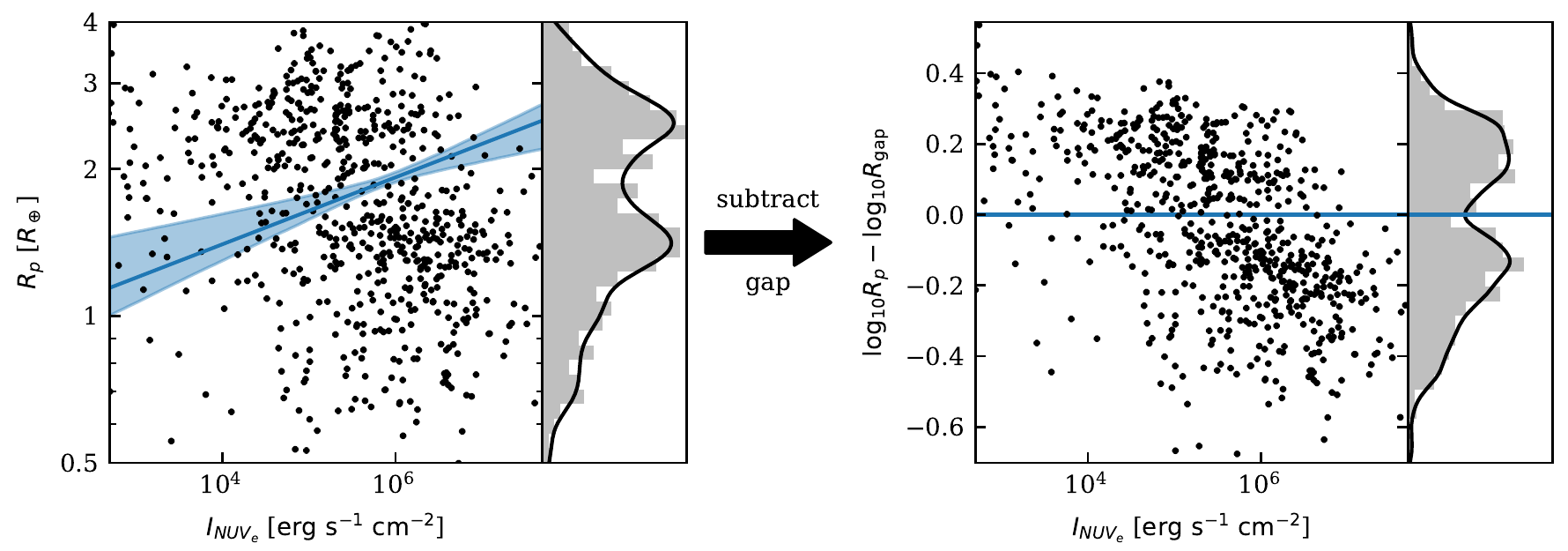}
\caption{Planetary radius gap in the $R_p$-\IxNUV\ (excess NUV planetary irradiation) plane.
\fitCaptionBoiler\
As with instellation, there is a strong positive relationship between the radius gap and \IxNUV.
\label{fig:xNUV}}
\end{figure*}

\begin{figure*}
\includegraphics{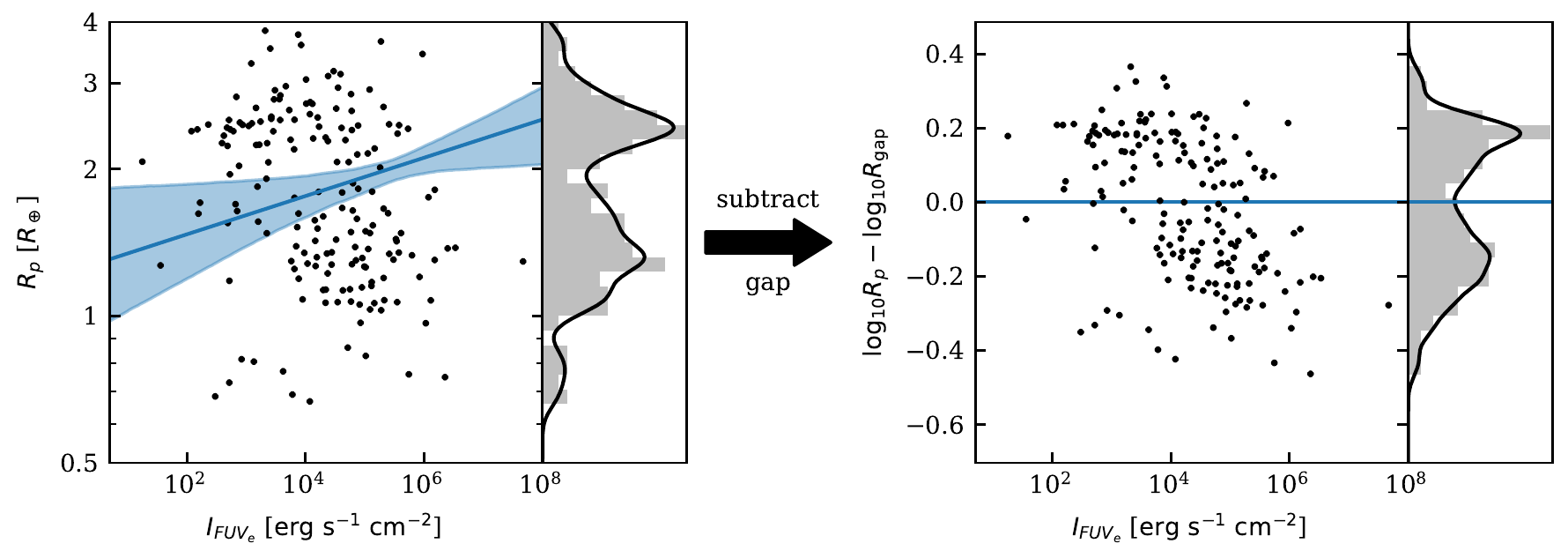}
\caption{Planetary radius gap in the $R_p$-\IxFUV\ (excess FUV planetary irradiation) plane.
\fitCaptionBoiler\
As with instellation, there is a strong positive relationship between the radius gap and \IxFUV.
\label{fig:xFUV}}
\end{figure*}

The gap's dependence on planetary irradiation is critical to this work.
Figure \ref{fig:S} shows the best fit for the gap in the $R_p$-$S$ plane, with a positive slope of $0.10^{+0.02}_{-0.04}$.\footnote{We quote uncertainties as the \nth{16}-\nth{84} (i.e., $1\sigma$) confidence intervals throughout this work.}
The radius gap has a positive trend with excess NUV irradiation and excess FUV irradiation of the planets, with slopes of $0.07^{+0.02}_{-0.03}$ (\IxNUV, Figure \ref{fig:xNUV}) and $0.04 \pm 0.03$ (\IxFUV, Figure \ref{fig:xFUV}).
Using the full NUV and FUV planetary irradiations without the photosphere subtraction described in Section \ref{sec:uv} produces $<1\sigma$ changes in slopes.
This indicates that the results are not sensitive to the details of the photosphere subtraction.

The relationships between the radius gap and \IxNUV\ and \IxFUV\ are not in themselves evidence for XUV evaporation.
Variations in \IxNUV\ and \IxFUV\ can be primarily attributed to variations in $S$.
Recall that  \IxNUV, \IxFUV, and $S$ are all fluxes measured \emph{at each planet}.
Therefore, they vary in unison in response to changes in semi-major axis and stellar luminosity.
Hence, even if the radius gap depends fundamentally only on $S$, as in the core-powered mass loss theory, the link between $S$,  \IxNUV, and \IxFUV\ mean the radius gap will trend with each of those values.

To disentangle XUV evaporation from core-powered mass loss, any population test must account for underlying (and, in the case of \IxNUV\ and \IxFUV, nearly 1:1) correlations with $S$.
The following subsections describe several such tests.
Because of the small size of the FUV sample, we restrict further discussion to the NUV sample only.
Numerical experiments confirm that each test operates as we intend (Appendix \ref{app:experiments}).

\subsection{No Relationship Between \lratioN\ and the Location of the Radius Gap after Accounting for Variations in Instellation}
\label{sec:splits}
Planets receiving the same instellation can have much different \IxNUV\ irradiations.
The difference arises from differences in \lratioN\ between stars.
A shift of the radius gap to greater radii for populations of planets orbiting higher \lratioN\ stars would be evidence for XUV evaporation.

We split the population of planets into thirds based on \lratioN\ to test for changes in the radius gap.
To isolate differences in the location of the gap, we fit the gap for each population under the constraint that their slopes be identical.

Figure \ref{fig:UVsplitfits} shows the gap fits (see Appendix \ref{app:splitfits} for individual fits).
Differences in the location of the gap between the populations with differing \lratioN\ are not statistically significant.
Allowing the slopes of each fit to vary independently does not change this result.

\begin{figure}
\includegraphics{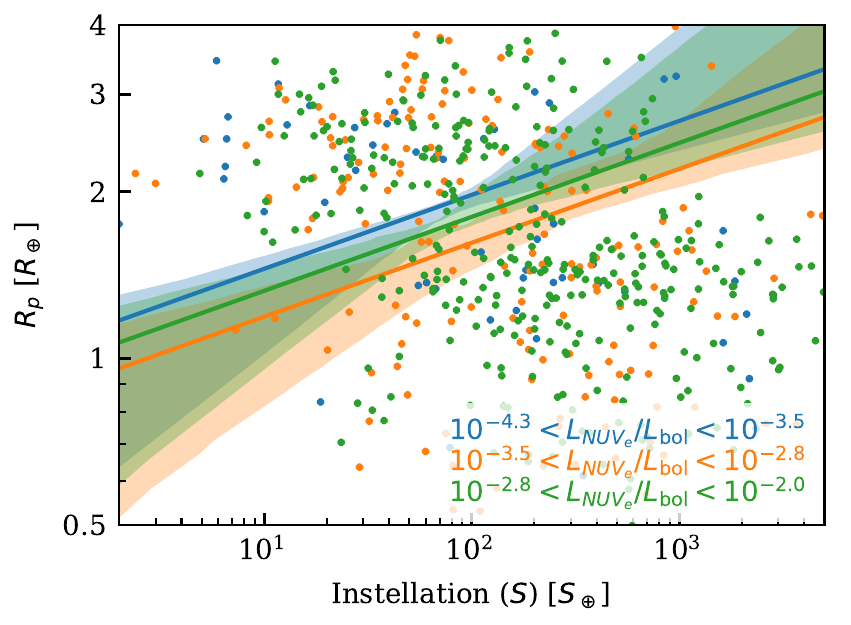}
\caption{Fits to the radius gap in the $R_p$-$S$ plane and 1$\sigma$ uncertainties for three planetary populations sorted by the \lratioN\ of their host stars.
No significant difference is present in the radius gap for these three populations.
\label{fig:UVsplitfits}}
\end{figure}

We carried out another test for a dependency on \lratioN\ that did not require binning.
For this, we detrended the planetary radii according to a fit to the gap in the $R_p$-$S$ plane by subtracting the $R_p$-$S$ fit in log space.
This yielded a radius gap that did not vary with $S$.
We then searched for a residual relationship between the detrended $R_p$ and \lratioN\ (see Appendix \ref{app:experiments}).

\begin{figure*}
\includegraphics{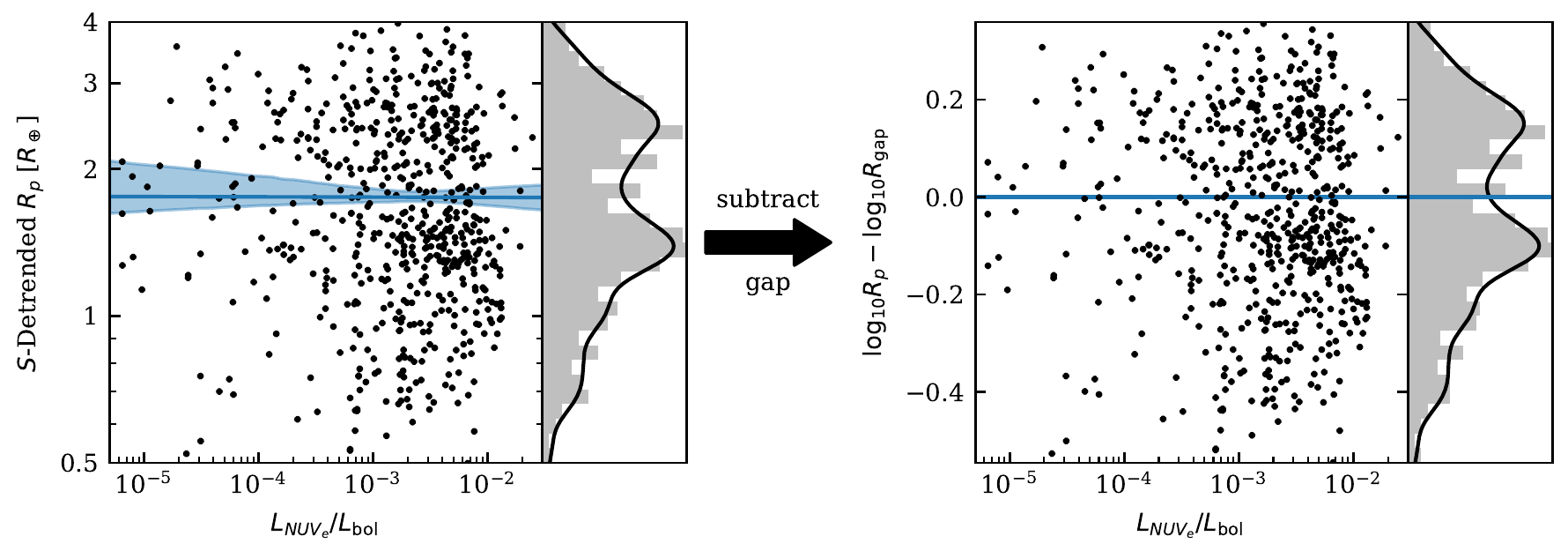}
\caption{Planetary radius gap in the $R_p$-\lratioN\ (fractional \xNUV\ luminosity of the star) plane after removing the dependence from the $R_p$-$S$ fit shown in Figure \ref{fig:S}.
\fitCaptionBoiler\
The slope of the radius gap is consistent with zero, implying that differences in the \xNUV\ portion of a planet's instellation are not associated with detectable changes in the radius gap.
\label{fig:best}}
\end{figure*}

The $S$-detrended radius gap shows no statistically-significant relationship with \lratioN.
The best fit slope is $0.00^{+0.01}_{-0.03}$ (Figure \ref{fig:best}).

\subsection{Gap Depth Is Not Greater when Using \IxNUV\ in Place of Instellation}
\label{sec:depth}
If XUV evaporation is the primary cause of the radius gap, the radius gap could be cleaner in the $R_p$-\IxNUV\ plane than the $R_p$-$S$ plane.
Variations in \lratioX\ will scatter points into the gap in the $R_p$-$S$ plane that would not be in the gap in the $R_p$-\IXUV\ plane.
If \IxNUV\ is an adequate proxy for early-life \IXUV, then the same will hold for it.
Measurement errors will also scatter points into the gap, potentially masking differences in its clarity.

\begin{figure*}
\includegraphics{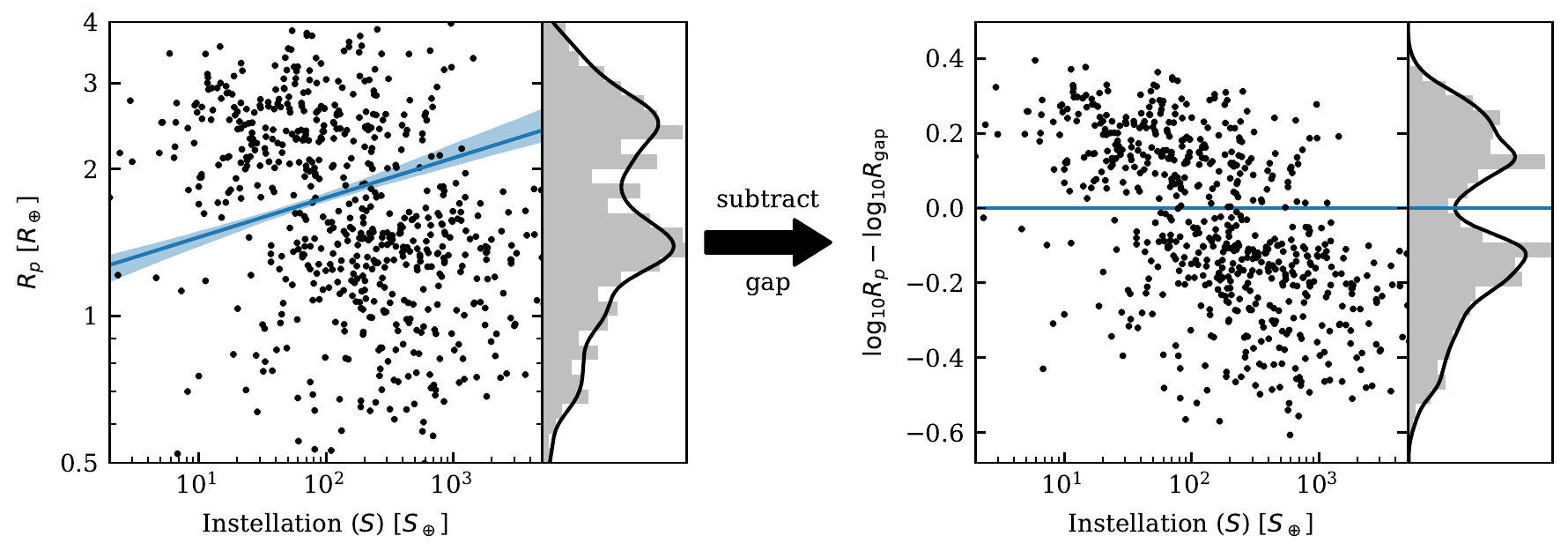}
\caption{Planetary radius gap in the $R_p$-$S$ plane using a planet sample identical to Figure \ref{fig:xNUVcompS}.
The purpose of this figure is to enable a comparison of the radius gap in the $R_p$-$S$ vs. $R_p$-\IxNUV\ planes using the same sample of 624 planets.
This excludes the 741 of 1365 planets in Figure \ref{fig:S} that have no \IxNUV\ measurement and the 73 of 697 planets in Figure \ref{fig:xNUV} that have no $S$ measurement. 
\fitCaptionBoiler\
The gap, the relative minimum in the right histogram, is deeper in this plane than the $R_p$-\IxNUV\ plane (Figure \ref{fig:xNUVcompS}).
\label{fig:ScompxNUV}}
\end{figure*}

\begin{figure*}
\includegraphics{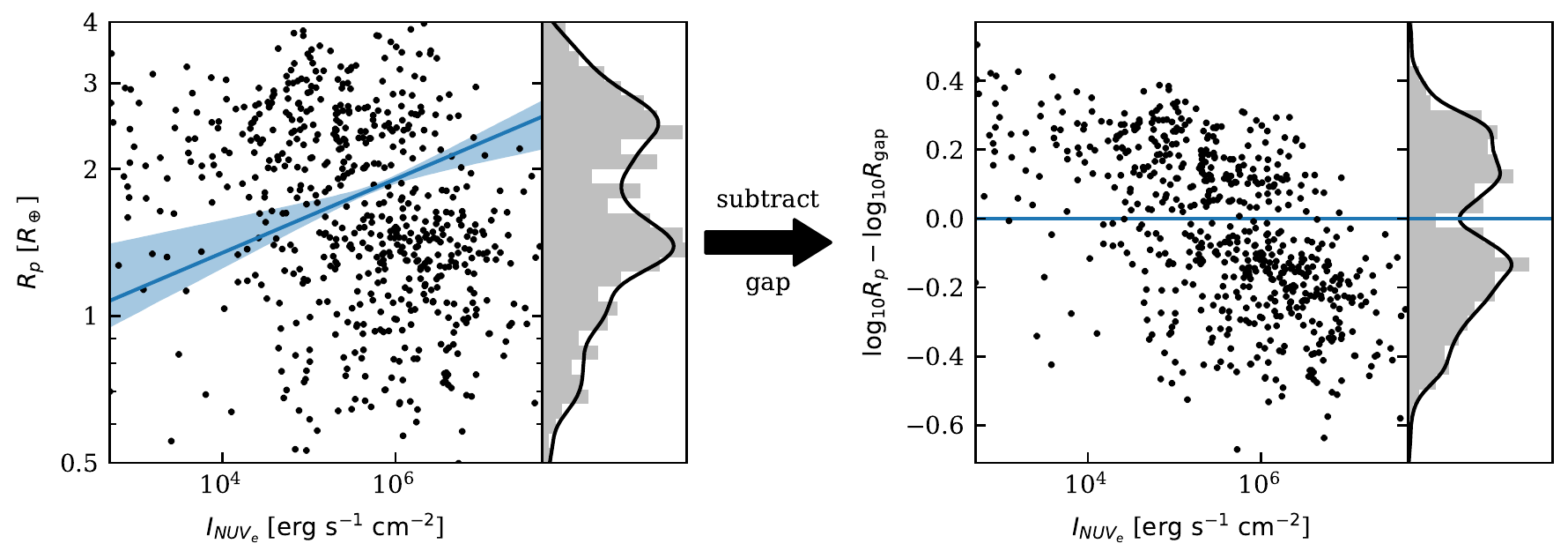}
\caption{Planetary radius gap in the $R_p$-\IxNUV\ plane using a planet sample identical to Figure \ref{fig:ScompxNUV}.
The purpose of this figure is to enable a comparison of the radius gap in the $R_p$-$S$ vs. $R_p$-\IxNUV\ planes using the same sample of 624 planets.
This excludes the 741 of 1365 planets in Figure \ref{fig:S} that have no \IxNUV\ measurement and the 73 of 697 planets in Figure \ref{fig:xNUV} that have no $S$ measurement. 
\fitCaptionBoiler\
The gap, the relative minimum in the right histogram, is not as deep in this plane as it is in the $R_p$-$S$ plane (Figure \ref{fig:ScompxNUV}).
\label{fig:xNUVcompS}}
\end{figure*}

We tested for differences in the clarity of the gap by measuring its depth. 
To ensure a consistent comparison, we used only planets with measurements of both $S$ and \IxNUV, the intersection of the samples in Figures \ref{fig:S} and \ref{fig:xNUV}.
We found that the gap is no deeper in the $R_p$-\IxNUV\ plane than in the $R_p$-$S$ plane (Figures \ref{fig:ScompxNUV} and \ref{fig:xNUVcompS}).

Positive results in Sections \ref{sec:splits} and \ref{sec:depth} would have been evidence for XUV evaporation.
However, the null results carry little weight.
If the present-day \xNUV\ emission of stars does not trace past differences in their XUV emission, XUV evaporation would produce no inherent radius gap -- \lratioN\ relationship.
Because the null results of the NUV analysis are uninformative, we will not discuss them further in this paper, instead focusing on the more informative results that rely on stellar mass.
However, the potential power of the tests in Sections \ref{sec:splits} is motivation to search for observational diagnostics that could trace the early XUV emission \textit{individual} stars observed late in life.

\subsection{No Relationship Between Host Star Mass and the Location of the Radius Gap after Accounting for Variations in Instellation}
\label{sec:mass}

\begin{figure}
\includegraphics{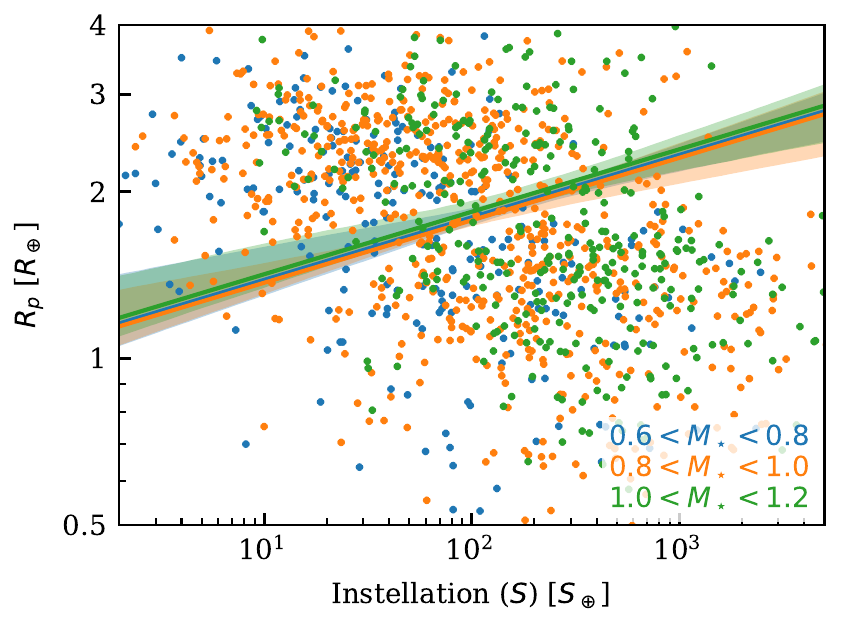}
\caption{Fits to the radius gap in the $R_p$-$S$ plane for three planetary populations sorted by the mass of their host stars.
No significant difference is present between these three populations with differing host star mass ranges. 
\label{fig:Msplitfits}}
\end{figure}

\begin{figure*}
\includegraphics{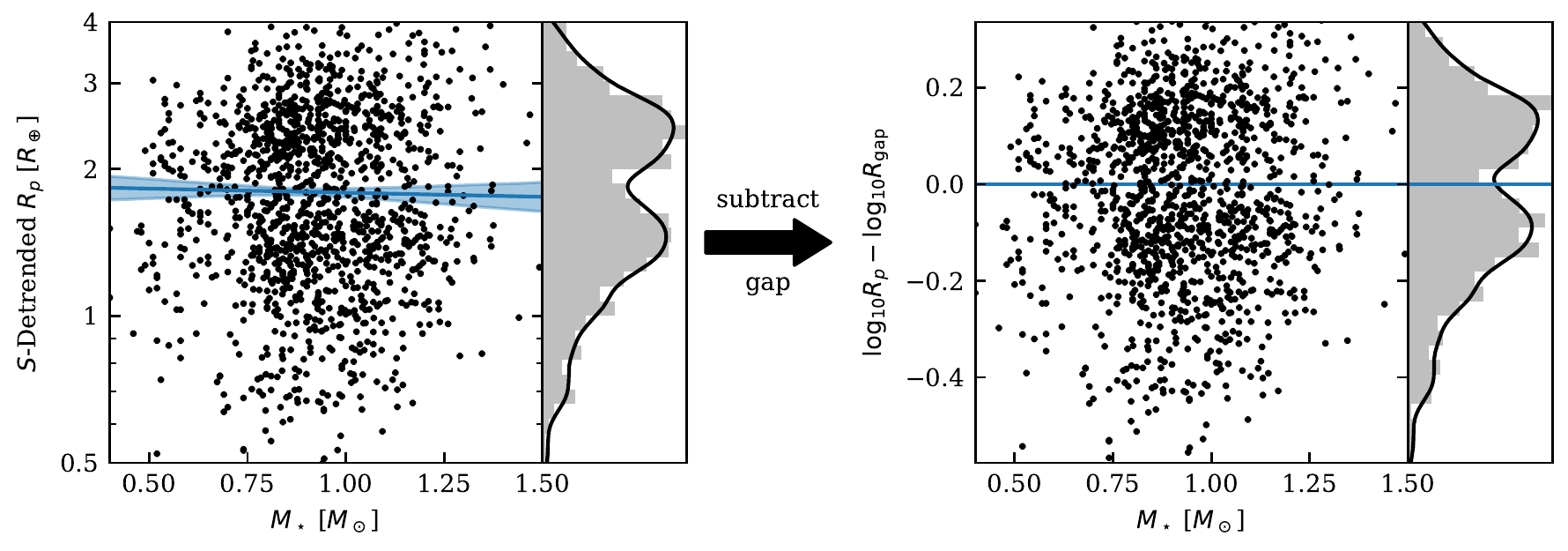}
\caption{Planetary radius gap in the $R_p$-$\Mm$ plane, with radii detrended according the underlying relationship between radius gap and instellation.
\fitCaptionBoiler\
After $S$-detrending, the slope of the radius gap is consistent with zero, implying that differences in $S$ can account for the relationship between radius gap and host star mass reported in previous works.
\label{fig:MS}}
\end{figure*}

We applied the same tests described in Section \ref{sec:splits}, but using host star mass in place of \lratioN.
The location of the gap does not vary for populations of planets orbiting stars in various mass bins (0.6-0.8, 0.8-1.0, and 1.0-1.2~$M_\odot$; Figure \ref{fig:Msplitfits}).
Similarly, after $S$-detrending, the radius gap shows no statistically significant trend with stellar mass (slope $-0.02\pm{0.05}$; Figure \ref{fig:MS}).
If we restrict the planet sample to that of \cite{fulton18} to ensure consistency in the measurement of system parameters, the slope becomes $0.00^{+0.07}_{-0.10}$.
The substantial increase in uncertainty results from the comparatively confined stellar mass range of the \cite{fulton18} sample.

We did not test for any change in the depth of the gap when plotted in the $R_p$-$\Mm$ plane versus the $R_p$-$S$ plane, analogous to that described in Section \ref{sec:depth}.
In the $R_p$-$\Mm$ plane, differences in the instellations of planets orbiting stars of similar mass will introduce large amounts of scatter, partially filling the gap.
Therefore, the gap should not be any deeper in the $R_p$-$\Mm$ plane than the $R_p$-$S$ plane.

\needspace{6em} \section{Discussion}
\label{sec:discuss}

XUV evaporation should imprint a stellar mass dependency on the exoplanet radius gap. 
As discussed in the introduction, because $\lratioXm \propto \Mm^{-2}$, lower-mass stars can strip atmospheres from larger planets at a given $S$.
Hence, after accounting for differences in $S$, the gap should move to greater radii for planetary populations orbiting lower-mass stars.
We do not observe this.
However, whether this result is significant depends on whether the trend predicted under XUV evaporation is strong enough to be detectable. 

To interpret the nondetection of a radius gap - stellar mass dependency, we simulated the atmospheric stripping of a synthetic planetary population.
For this, we used the minimum analytical XUV evaporation model described in \cite{owen17} (see also \citealt{wu19} and \citealt{owen19x}).
This model begins with a population of planets Rayleigh distributed in core mass ($\sigma_{M_c} = 3 \mMearth$) and with a uniform distribution of atmospheric masses between 1\% and 30\%.
We initialized the population with the host star masses and planetary orbital periods of the real-world sample.
We assumed the same XUV emission model as in \cite{owen17},  $L_{XUV} = 10^{-3.5}~\mLsun~M_\star/\mMsun$ for 100~Myr followed by a $t^{-1.5}$ decline.
The model steps through time, solving for the radius of the optically-thick atmosphere of a given planet, determining the amount of XUV energy that atmosphere absorbs, and computing the amount of atmospheric mass the absorbed energy liberates, assuming 10\%  efficiency.

\begin{figure}
\centering
\includegraphics{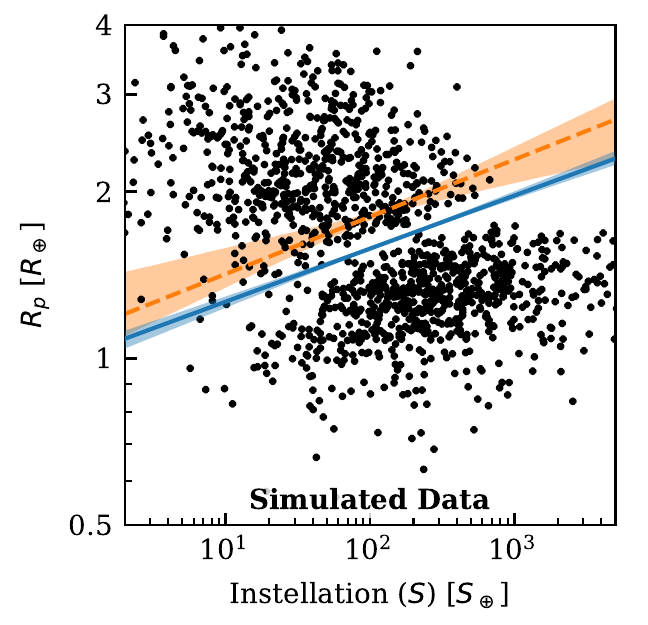}
\caption{The instellation dependency of a simulated radius gap generated by XUV evaporation.
The simulated population in this figure can be directly compared to the real planetary population in the left panel of Figure \ref{fig:S}.
The radius gap has a slope of $0.096_{-0.008}^{+0.009}$ (blue line) in agreement with the slope of the radius gap in the actual planet population (orange line) of $0.10_{-0.04}^{+0.02}$.
See text for discussion of the offset between the simulated and actual radius gaps.
\label{fig:simfit_S_control}}
\end{figure}

The simulation produces a radius gap similar to that of the actual planetary population.
Figure \ref{fig:simfit_S_control} plots the simulated planetary population in the $R_p$-$S$ plane, with a fit to its gap and the fit to the gap in the real planet population overplotted.
The slopes of the gaps are identical.
However, they differ notably in location.
The same discrepancy is not present in \cite{owen17}; their simulated gap's location matches reality.
The cause of the difference is the assumed host star masses.
\cite{owen17} initialize their simulation with stars Gaussian distributed around a median mass of 1.3~\Msun\ with $\sigma=0.3$~\Msun.
We initialize our simulation with the actual host star masses, median mass 0.96~\Msun.
The greater luminosities of the stars in the \cite{owen17} simulation yield planet instellations roughly an order of magnitude larger than those of the actual planet sample.
At those instellations, the location of the simulated gap is at 1.8~\Rearth, as observed.
However, XUV evaporation simulations of planetary populations can be tweaked to better match reality.
\cite{wu19} reproduces the observed gap with the same model using actual host star masses.
One possible factor enabling the better agreement between the simulated and real radius gap in \cite{wu19} is that \cite{wu19} prescribes a variable mass loss efficiency.

\begin{figure}
\centering
\includegraphics{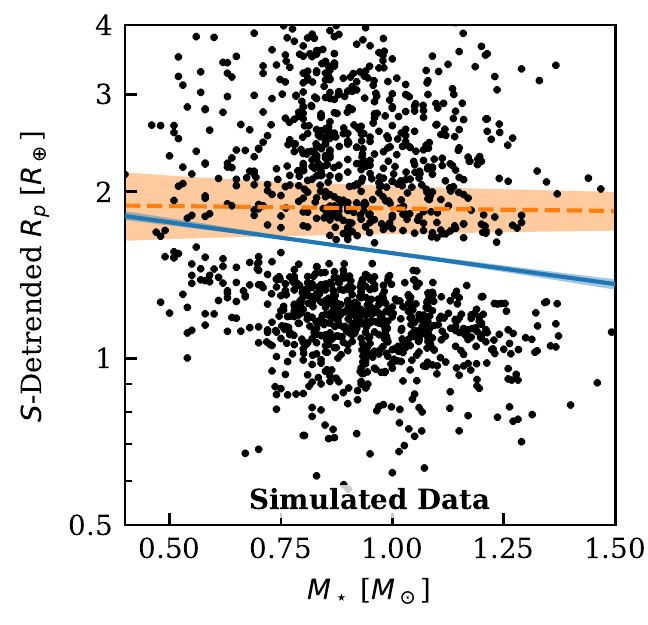}
\caption{The stellar mass dependency of a simulated exoplanet radius gap generated by XUV evaporation.
The slope of the simulated radius gap, $-0.11_{-0.02}^{+0.01}$ (blue line), is steeper than the actual radius gap, $-0.02\pm0.05$ (orange line).
However, adding measurement error and other complications eliminates this difference (see Figure \ref{fig:simfit_M_S_complicated}).
\label{fig:simfit_M_S_control}}
\end{figure}

We tested the simulated planetary population for a dependency between the radius gap and stellar mass after $S$-detrending, just as we did with the actual planet population.
A stellar mass dependency is clear in the simulated sample, with a radius gap slope of $-0.11_{-0.02}^{+0.01}$ (Figure \ref{fig:simfit_M_S_control}).
This confirms that, under XUV evaporation, the radius gap should be inversely related to stellar mass once the underlying dependency on instellation has been removed.

The simulation shown in Figures \ref{fig:simfit_S_control} and \ref{fig:simfit_M_S_control} is an ideal case, assuming no measurement errors, no star-to-star variations in UV activity, and uniform system ages.
In reality, a bevvy of complications blur the radius gap and its dependencies on system parameters:
\begin{itemize}
    \item Adding measurement uncertainty to the planetary radii, stellar masses, and planetary instellations increases the error on the recovered slopes.
    At 10\% radius error, all other values assumed exact, the significance of the simulated trend with stellar mass is reduced to $3\sigma$.
    At 20\% radius error, most fits to bootstrapped samples do not converge.
    The effects of errors on stellar mass and planetary instellation are less severe.
    For example, with 20\% errors on stellar mass, the significance of the simulated trend becomes $6\sigma$.
    \item Evolving the planets to the catalog ages of the systems (bootstrapping the $\sim$10\%\ of systems without catalog age values) has little effect.
    This is because system ages are almost all $> 1$~Gyr, whereas most atmospheric stripping occurs within the first 100~Myr.
    \item Incorporating an updated stellar XUV evolution model weakens the radius gap slope to $-0.07\pm0.01$ in the $S$-detrended $R_p$-$S$ plane.
    The updated model is a simplification of the results presented in \cite{mcdonald19}, with $\lratioXm \propto 10^{-3.7} (\Mm/M_\odot)^{-2.2}$ for 100~Myr followed by a $t^{-1.2}$ decline.
    \item Varying the stellar XUV flux by 0.5 dex to mimic the observed variability in X-ray and UV emission between stars (see Introduction) weakens the gap slope to $-0.06\pm0.02$ in the $S$-detrended $R_p$-$S$ plane.
    It is notable that the gap itself, not just its stellar mass dependency, remains detectable in spite of this scatter, meaning that the XUV evaporation model is robust to variability in the XUV evolution of individual stars.
    However, possible order of magnitude differences in saturation timescale between individual stars could make this 0.5~dex prescription optimistic (see Section \ref{sec:intro}).
\end{itemize}

\begin{figure}
\centering
\includegraphics{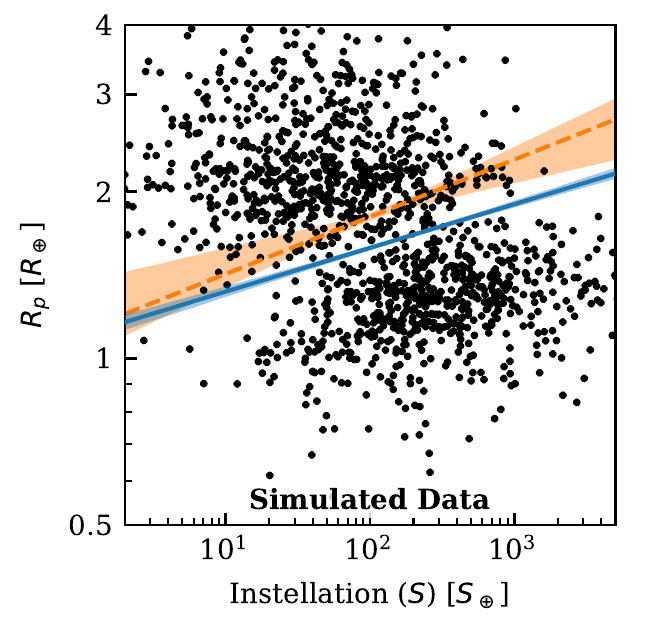}
\caption{The instellation dependency of a simulated radius gap generated by XUV evaporation, with added measurement error and other complications (see text).
The radius gap has a slope of $0.079\pm0.006$ (blue line), consistent with the observed slope of $0.10_{-0.04}^{+0.02}$ (orange line).
\label{fig:simfit_S_complicated}}
\end{figure}

\begin{figure}
\centering
\includegraphics{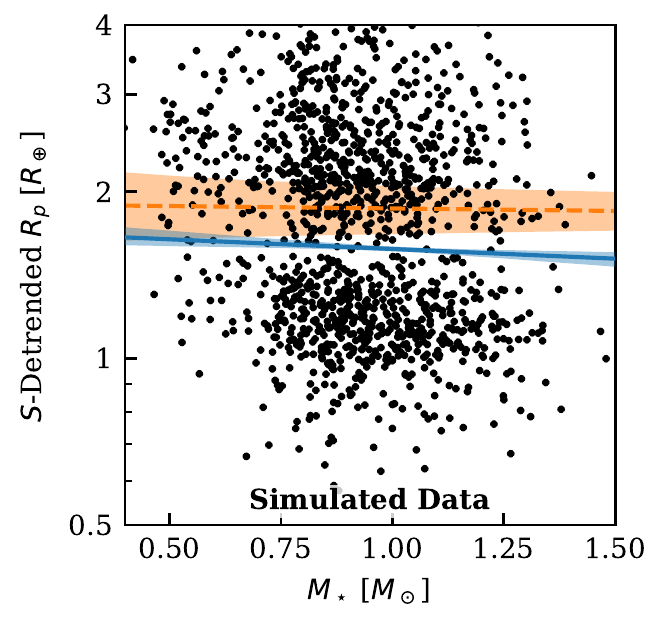}
\caption{The stellar mass dependency of a simulated radius gap generated by XUV evaporation, with added measurement error and other complications (see text).
The radius gap has a slope of $-0.035_{-0.03}^{+0.02}$ (blue line), in agreement with the actual radius gap slope of $-0.02\pm0.05$ (orange line).
Including realistic treatments of stellar XUV evolution and measurement errors makes the predicted radius gap trend with \M\ under XUV evaporation no longer clearly detectable.
\label{fig:simfit_M_S_complicated}}
\end{figure}

To create the most realistic simulation possible, we combined the above modifications into a single simulation.
We injected measurement errors at the median levels of the actual population: 5\% in $R_p$, 3\% in $M_\star$, 7\% in $S$.
The resulting population (Figures \ref{fig:simfit_S_complicated} and \ref{fig:simfit_M_S_complicated}) does not have a detectable radius gap - stellar mass trend.
Its radius gap has a slope of $-0.035_{-0.03}^{+0.02}$ in the $S$-detrended $R_p$-$S$ plane, consistent with the $-0.02\pm0.05$ radius gap slope of the actual population.
Hence, our tests ultimately do not exclude XUV evaporation as the primary cause of the radius gap.
Core-powered mass loss also remains viable as a primary cause of the radius gap because it involves no mechanism that would produce a dependency on host star mass after $S$-detrending.

The line of reasoning we have developed could plausibly discriminate between these theories with an expanded exoplanet sample.
As the exoplanet sample increases, particularly as the diversity of host-star masses grows, the sensitivity of the test for a  radius gap - stellar mass trend improves.
Adding a sample of 2000 exoplanets with host star masses drawn from a uniform distribution across 0.3~-~2.0~\Msun\ to the existing sample yields a $>3\sigma$ trend in our simulation. 

The XUV evaporation and core-powered mass loss theories are not mutually exclusive.
It is well-established that XUV radiation is what heats the upper atmospheres of solar-system planets, resulting in slow Jean's escape of light species (e.g., \citealt{hunten93}).
Both XUV evaporation and core-powered mass loss likely operate with varying efficacy over the course of every planet's life.
However,  our tests with stellar mass hint that XUV evaporation might not be the dominant cause of the radius gap.

\subsection{Previous Results of the Radius Gap's Relationship with Stellar Mass}
Four previous studies have explicitly examined the relationship between stellar mass or equivalent properties and the radius of the gap.
\cite{zeng17} split up all confirmed and candidate planetary systems in the NASA Exoplanet Catalog into late-M, early-M, K, and G star groups.
They noted an increase in the radius of the gap toward earlier host star types.
They posit that this relationship is a result of higher planetary masses around stars of earlier type.
A more in-depth analysis by \cite{wu19} of planetary radii versus host star mass reaches the same conclusion, using the model of \cite{owen17} to simulate atmospheric stripping for a population of planets where the mass of the planetary cores depends on the mass of the host stars.
They find that they can reproduce several features of the planetary population that vary with stellar mass, including the radius where the super Earth population peaks, the radius where the gap is deepest, and the radius where the sub Neptune population peaks.

We offer an alternative interpretation of these results.
A direct relationship between the mass of planetary cores and the mass of the host stars will, undoubtedly, influence the demographics of the planetary population.
The relationship would cause the typical radius of sub Neptunes to increase with stellar mass, all else being equal.
This, in turn, would leave fewer planets small enough to have their atmospheres stripped, depleting the population of super Earths as stellar mass increases.
However, the location of the radius gap would not change.
Under the XUV evaporation model, atmospheric stripping is a simple energy balance between a planet's gravity and its XUV irradiation.
If the XUV irradiation is kept constant, the transition between planets that can retain a primordial atmosphere and planets stripped of them will occur at the same core mass and radius.
If the distribution of core masses (and radii) is changed, the populations on either side of the gap will adjust accordingly, but the radius gap itself will not move.
Only second order shifts might result from changes to the number density of sub Neptunes and super Earths.

\begin{figure}
\centering
\includegraphics{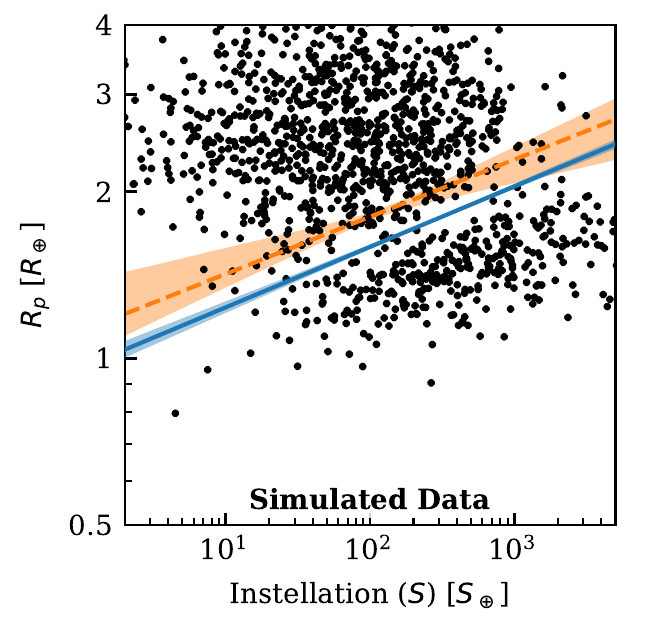}
\caption{The instellation dependency of a simulated radius gap generated by XUV evaporation.
In this simulation, planetary core masses depend linearly on the host star mass as per \cite{wu19} (see text).
The radius gap formed in this simulation has a slope of 0.11 with respect to $\log_{10}S$ and crosses 1.59~\Rearth\ at $S = 100\ S_\oplus$ (blue line), compared to 0.10 and 1.58~\Rearth\ for the original simulation  (Figure \ref{fig:simfit_S_control}).
The orange line shows the radius gap fit to the actual exoplanet population.
This demonstrates that the radius gap does not significantly depend on the initial planetary mass distribution in an XUV evaporation simulation.
\label{fig:simfit_S_star_planet_corr}}
\end{figure}

\begin{figure}
\centering
\includegraphics{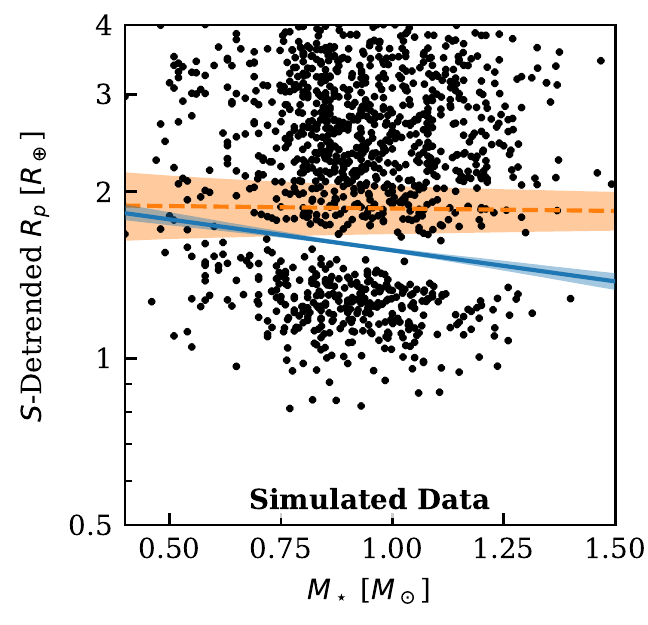}
\caption{The stellar mass dependency of a simulated radius gap generated by XUV evaporation.
In this simulation, planetary core masses depend linearly on the host star mass as per \cite{wu19} (see text).
The radius gap formed in this simulation has a slope of -0.1120 with respect to \M\ (after $S$ detrending) and crosses 1.57~\Rearth\ at $\Mm = 1\ M_\odot$, compared to -0.1117 and 1.55~\Rearth\ for the original simulation (Figure \ref{fig:simfit_M_S_control}; values quoted with excess significant digits to illustrate the smallness of the difference between simulations).
The orange line shows the radius gap fit to the actual exoplanet population.
This demonstrates that a planet-star mass correlation has minimal impact on the inverse relationship between the radius gap and stellar mass after $S$-detrending.
\label{fig:simfit_M_S_star_planet_corr}}
\end{figure}

To validate the reasoning that the radius gap should not depend on planetary masses, we simulated atmospheric stripping using a distribution of planet masses dependent on  host star mass.
Specifically, we drew core masses from a normal distribution in log space with a mean mass of $M_\mathrm{core} = 8\ M_\oplus\ (\Mm/M_\odot)$ and a standard deviation of 0.3 dex, the preferred model of \cite{wu19}.
To isolate the effect of the assumed mass distribution, we kept all other parameters of the model identical to the simulation described in Section \ref{sec:discuss}.
Figures \ref{fig:simfit_S_star_planet_corr} and \ref{fig:simfit_M_S_star_planet_corr} show the results.

The radius gap fits to the population synthesized with a star-planet mass correlation are nearly identical to the population synthesized with no such correlation.
From this, we conclude that the radius gap itself is insensitive to the initial planetary mass distribution, as expected.
However, a relationship between host star and planet masses could nonetheless exist.
The shape of the super Earth and sub Neptune distributions are sensitive to the initial planetary masses, and \cite{wu19} shows that a planet-star mass correlation reproduces them well.

Returning to the results of other works addressing the mass-dependency of the radius gap, \cite{fulton18} found a positive relationship between the radius gap and host star mass in the \textit{CKS-Gaia} sample.
They conclude this relationship supports the XUV evaporation model, arguing that the population of sub-Neptunes should shift to lower instellations for lower-mass stars because lower-mass stars emit a larger fraction of their lifetime radiation in the XUV range.
However, shifting the sub-Neptune population to lower instellations, without shifting the super-Earth population,  would widen the gap in the $R_p$-$S$ plane and move the radius at the center of the gap to a larger value.
This would cause the radius of the gap to increase with decreasing stellar mass, contrary to the observed trend.

\cite{gupta19b} simulated populations of planets subjected to the core-powered mass loss mechanism to compare to observations.
They were able to recover the observed dependency on stellar mass when using the same distributions of stellar masses, luminosities, and planetary orbital periods as the observed sample.
They attribute the radius gap -- stellar mass trend to the strong relationship between stellar mass and luminosity that could bias planets orbiting more massive stars toward greater instellations. 
Our independent recognition of the importance of instellation in possible stellar mass trends is what motivated the $S$-detrending of the present work. 

\begin{figure}
\includegraphics{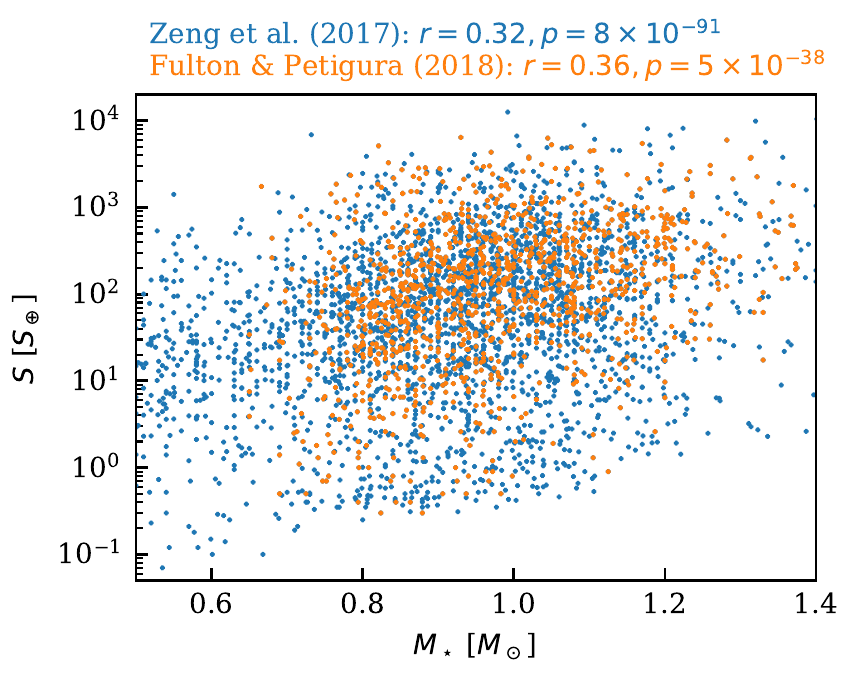}
\caption{The correlation between stellar mass and planetary instellation in the \textit{Kepler} confirmed and candidate planet sample with $R_p < 4\ R_\oplus$ (similar to \citealt{zeng17}) and the \cite{fulton18} planet sample.
This correlation can explain the dependency between the radius gap and stellar mass reported in previous works.
\label{fig:MScorr}}
\end{figure}

Based on our analyses, we conclude the simplest explanation for the results from \cite{zeng17}, \cite{fulton18}, and \cite{wu19} is the underlying correlation between planet instellation and stellar mass first recognized by \cite{gupta19b}.
In Figure \ref{fig:MScorr}, we plot this correlation for a sample of planets similar to the \cite{zeng17} and \cite{fulton18} samples.
Despite the large scatter, a trend is clearly present. 
For the \cite{zeng17} planets, a Spearman rank-order correlation test yields an $r$ of 0.32 and a p-value of $10^{-90}$ and, for the \cite{fulton18} planets, an $r$ of 0.36 and a p-value of $10^{-38}$.
These correlations are of high confidence.
Classifying the sample by stellar type according to the stellar radius cuts in \cite{zeng17}, we find median instellations of 6.3 $S_\oplus$ for planets orbiting late Ms ($N=24$), 15 $S_\oplus$ for early Ms ($N=242$), 62 $S_\oplus$ for Ks ($N=1568$), and 100 $S_\oplus$ for Gs ($N=798$).
\cite{wu19} adopted the orbital periods and stellar masses of an actual exoplanet population as simulation input, thereby implicitly incorporating a correlation between stellar mass and planetary instellation in their XUV evaporation simulations.

The larger instellations of known planets orbiting more massive stars means atmospheres can be stripped from larger planets according to both the XUV evaporation and core-powered mass loss models.
This provides a simple explanation for the dependency of the gap on stellar type observed by \cite{zeng17} and \cite{fulton18} and reproduced in simulations by \cite{wu19} and \cite{gupta19b}, as well as the results of this paper.
Our analysis has shown that once instellation has been accounted for, no detectable dependency on stellar mass remains (Figures \ref{fig:Msplitfits} and \ref{fig:MS}).

\subsection{A Related Study of the sub-Jovian Desert}
\cite{mcdonald19} recently investigated a different feature in the planetary radius-instellation plane from the perspective of stellar mass and activity, the sub-Jovian desert.
The sub-Jovian desert refers to a lack of planets with radii in the sub-Neptune to Jupiter range and orbital periods of a few hours to a few days.
\cite{mcdonald19} constructed a sophisticated model to predict the lifetime XUV irradiation of these planets based on host-star mass.
They find that plotting planetary radii versus predicted lifetime-XUV exposure sharpened the lower edge of the sub-Jovian desert, and the planetary radius of the edge increased with decreasing lifetime XUV exposure, as expected.
They take this as evidence for XUV evaporation as the cause of the sub-Jovian desert.
This contrasts with our finding of no detectable dependence of the radius gap on stellar mass.

\section{Summary and Conclusions}
\label{sec:summary}
We have analyzed the relationship between the observed planetary radius gap and two candidate proxies for the early-life XUV-radiation of stars: their present-day, excess NUV flux attributable to stellar activity (\xNUV) and their mass.
Our aim was to search for empirical evidence of XUV evaporation in the population of known planets.

The radius of the gap shows no dependence on the fractional \xNUV\ emission of the stellar hosts (\lratioN).
This could simply indicate the present-day activity level of stars is a poor correlate for their early-life activity, and is not evidence against XUV evaporation as the cause of the radius gap.
As a product of this work, we make available a catalog of NUV and FUV measurements or upper limits from the \textit{GALEX} observatory archive for nearly all stars presently known to host planets.

The radius of the gap also shows no statistically significant trend with stellar mass after accounting for differences in planetary instellations.
This accounting is critical to remove the effects of an underlying correlation between instellation and host star mass in the exoplanet sample.
Tests with simulated planetary populations subjected to XUV evaporation suggest that an inverse trend between the radius gap and host star mass is detectable with the current exoplanet population under ideal circumstances.
However, adding plausible scatter in the XUV emission of individual host stars and realistic measurement errors, along with several less impactful factors, weakens the predicted stellar mass dependency and increases measurement uncertainties to the point that the simulated and observed radius gap trends are consistent within 1$\sigma$.

We found independently that the positive trend between the radius gap and stellar mass reported in several previous works can be explained by the underlying planetary instellation - stellar mass correlation, confirming the interpretation of \cite{gupta19b}.
Further, using XUV evaporation simulations, we showed that assuming a  relationship between stellar and planetary masses does not appreciably change the radius gap trends retrieved by our analysis.

By detecting and accurately characterizing more exoplanets orbiting M and K stars to better sample disparate XUV emission histories, this line of evidence could eventually discriminate between core-powered mass loss and XUV evaporation as dominant causes of the radius gap.
In addition, more precise empirical constraints on stellar XUV evolution coupled with detailed modeling of planetary atmospheric loss are critical to increasing confidence in the predictions of the radius gap's dependence on stellar mass under the XUV evaporation theory.

\acknowledgments

R.O.P.L. and E.S. gratefully acknowledge support from NASA \textit{HST} Grant  HST-GO-14784.001-A for this work.
We sincerely thank James Owen for providing us with the code from \cite{owen19x}.
We wish to express deep gratitude for the monumental effort of the teams carrying out the Gaia, Kepler, GALEX, CKS, asteroseismology and smaller planetary and stellar surveys that make population studies of this kind possible.
We thank Akash Gupta, Jessica Roberts, Mike Line, Tad Komacek, Patrick Young, and Aishwara Iyer for useful discussions.
This research has made use of NASA's Astrophysics Data System and the NASA Exoplanet Archive, which is operated by the California Institute of Technology, under contract with the National Aeronautics and Space Administration under the Exoplanet Exploration Program.
This work is based on observations made with the NASA Galaxy Evolution Explorer.
GALEX was operated for NASA by the California Institute of Technology under NASA contract NAS5-98034.
We thank Megan Bedell for making a publicly available \textit{Gaia-Kepler} cross match catalog.
This work has made use of data from the European Space Agency (ESA) mission Gaia (https://www.cosmos.esa.int/gaia), processed by the Gaia Data Processing and Analysis Consortium (DPAC, \url{https://www.cosmos.esa.int/web/gaia/dpac/consortium}). Funding for the DPAC has been provided by national institutions, in particular the institutions participating in the Gaia Multilateral Agreement.

This work is dedicated to (ultra)Violet Schneider, born 2018 August 2, whose auspicious name naturally predisposes her to great feats of UV science.


\vspace{5mm}
\facilities{GALEX, NASA Exoplanet Archive}


\software{astropy \citep{astropy13}}.

\newpage

\appendix

\section{Fitting the Gap}
\label{app:fits}
Fitting a model to the relationship between the location of the radius gap and some independent variable such as orbital period presents an unusual challenge.
It requires selecting model parameters that result in, loosely speaking, the fewest data points possible near the model, rather than matching data to the model.
Ideally, a model of the data would directly predict the planetary populations above and below the gap, thus fitting for the observed planets rather than the gap.
However, this would require modeling the intrinsic planetary population, modeling the evolution of that population, and knowing and combining the detection completeness of each survey that contributed planets to the catalog.
This is a daunting level of complexity with a great deal of room for systematic errors.
Instead, it is more straightforward to simply fit the gap.

Several strategies have been employed in other works for this purpose.
These all fit power-law relationships between the radius gap and the independent variable of choice, i.e. a linear model in log-log space.
The techniques of which we are aware are:
\begin{enumerate}
    \item Finding the peaks of the planet distributions above and below the gap, then taking the slope of the gap model to be that of a line orthogonal to the line connecting those peaks \citep{macdonald19}.
    \item Computing the likelihood of the data, then finding the local \emph{minimum}, rather than maximum, likelihood that places the model line within the gap \citep{eylen18}.
    \item Binning the planets according to the independent variable (e.g., orbital period) and fitting a line to features of the planetary radii in those bins (e.g., the maximum planet radius below the gap, \citealt{eylen18}; or the radius of minimum  planet number density, \citealt{martinez19}).
    \item Using a support vector machine, a machine learning method, to find the line that optimally separates the planets above and below the gap \citep{eylen18}.
\end{enumerate}

The three different techniques used by \cite{eylen18} produced consistent results within 1$\sigma$ uncertainties.
The \cite{macdonald19} approach has the advantage of requiring little human intervention in comparison to the methods used by \cite{eylen18} and the technique presented here.
It is in principle possible to use unsupervised machine learning to both identify the two populations of planets, as in \citealt{macdonald19}, and then fit for a line optimally separating those populations, as in \citealt{eylen18}.
However, we feel that it is reasonable to use human intelligence to guide a numerical optimizer that finds best-fit parameters for a gap model.

\subsection{The ``Deepest Gap'' Method }
In this work, we have employed a custom method to fit the radius gap.
Our motivation to develop a custom algorithm, versus using those already employed in the radius gap literature, stemmed from a desire for three qualities:
\begin{enumerate}
    \item that it use all the data available, in contrast to methods (3) and (4) of the previous section,
    \item that it be reasonably insensitive to the effects of varying detection completeness, which might not be the case for method (1) of the previous section,
    \item that it mimic the human intuition we use in visually identifying the gap.
\end{enumerate}
These goals, in particular goal (3), led us to develop a method that finds the line along which the gap is deepest.

To do this, the algorithm subtracts a guess at the line describing the gap from the log of the planetary radii.
Then, it uses kernel density estimation (KDE) to estimate the number density of planets (i.e., number of planets per increment in log radius) at the gap using the transformed data (e.g., Figure \ref{fig:S}).
For models that better align the gap, the density of planets in the gap is lower.
The best-fit is considered to have been reached when the depth of the gap is maximized (the number density of planets at the line of the gap is minimized).
A Python implementation of this gap fitting scheme is available on GitHub.\footnote{\url{https://github.com/parkus/gapfit}}

We use a linear model of the gap defined by a slope, $m$; a pivot point in the independent variable, usually the log of some property such as orbital period, $x_0$; and the log radius at mid-gap at that pivot point, $\lrz$;
\begin{equation}
    \lrgap = m (x - x_0) + \lrz.
\end{equation}
The free parameters of the fit are $m$ and $\lrz$; $x_0$ is fixed.
Choosing an $x_0$ near the center of the data is helpful because it minimizes correlations between $m$ and $\lrz$ in the fit.
Theoretically, the choice of $x_0$ has no effect on the fit itself, though, in practice, very strong correlations in fit parameters can sometimes pose a challenge to numerical optimizers.
In principle, this fitting scheme could be used with more complicated functional models of the radius gap.

An important nuance is that the spread of the data in log radius after the gap line is subtracted will differ from the spread before subtraction, or, more importantly, relative to the spread when using a different guess at the gap line.
Hence, after the transformation, we normalize the data to unit variance in the gap-subtracted log radii.

Next, we apply KDE to the gap-subtracted, variance-normalized log radii using a kernel width of 0.15 to estimate the density of planets at a gap-subtracted log radius of 0.
Using KDE to estimate the density in the gap is a subjective choice.
It could be done using a histogram or any arbitrary technique.
We chose KDE because it incorporates all of the data and avoids the potential step function in density that would result from histogramming when points are shifted in and out of bins as the model parameters are varied.
We chose a kernel width of 0.15 because it generally produced a clean bimodal distribution in planet density where the gap was apparent.
Values a little lower (e.g., 0.1) generally yielded curves with $>2$ modes.
Larger values would progressively blur the gap until it disappeared and the curve became unimodal around a kernel width of 0.4.
The best fit values were mildly sensitive to the choice of kernel width, but changes in the fit values were below 1$\sigma$ uncertainties for kernel widths from 0.1 to 0.4 and there was no clear trend between the kernel width used and the slope of the best-fit line. 

The density of the planets at the gap line was taken as the cost function for the fit and supplied to a numerical minimizer that found optimal values of $m$ and $\lrz$.
A challenge of this scheme is that the gap represents only a local minimum in planet density.
Hence, if the numerical minimizer takes a step large enough to leave the gap, it will simply begin moving the best-fit line further and further from the overall cluster of planets and will not converge.
We eliminated this problem by constraining the fit to reasonable values of $\lrz$ -- essentially placing a prior on $\lrz$ of (0.15, 0.35), equivalent to $R_0$ in the range (1.4, 2.2).
The data were bootstrapped to estimate realistic uncertainties on the fit.

We tested our fitting method on the $\lr$ versus $\log_{10} P$ data for the asteroseismic sample of \cite{eylen18} and retrieved $m=-0.10_{-0.04}^{+0.02}$.
In comparison, the three techniques tested by \cite{eylen18} yielded $-0.13^{+0.04}_{-0.05}$ (minimum likelihood fit), $-0.05^{+0.01}_{-0.03}$ (fit to binned maximum planet radii), and $-0.09^{+0.02}_{-0.04}$ (support vectors, chosen as their quoted value).
\cite{martinez19} and \cite{macdonald19} found slopes of  $-0.11\pm0.02$ and $-0.32^{+0.09}_{-0.12}$.
The \cite{macdonald19} is an outlier because they fit the peaks of the populations above and below the gap rather than the gap itself.

In our fits to the full sample, we found the retrieved slope was sensitive to the bounds of the planet sample used.
The slope generally stayed within 1$\sigma$ of the slope from the full sample given the increase in uncertainty associated with confining the sample.
Progressively removing planets from the extremes of the period distribution had the greatest effect, since this restricts the lever arm of the fit.
Restricting the sample to a period range of (5, 15) d yielded a sample size half of the original but yielded a slope of $-0.02^{+0.05}_{-0.04}$.
Below, we list the results from several fits with varying bounds, given as period range: sample size, $\lrz$, $m$.
We have averaged asymmetric errors.
\begin{itemize}
    \item (0, 100): 1548, 0.249 $\pm$ 0.007, -0.08 $\pm$ 0.01
    \item (0, 10): 861, 0.25 $\pm$ 0.01, -0.10 $\pm$ 0.03
    \item (10, 100): 687, 0.25 $\pm$ 0.03, -0.09 $\pm$ 0.04
    \item (5, 50): 993, 0.24 $\pm$ 0.01, -0.05 $\pm$ 0.03
    \item (10, 30): 499, 0.24 $\pm$ 0.03, -0.05 $\pm$ 0.10
    \item (5, 15): 751, 0.24 $\pm$ 0.01, -0.02 $\pm$ 0.04
    \item (0, 100), every third planet: 516, 0.25 $\pm$ 0.01, -0.07 $\pm$ 0.02
\end{itemize}

\section{A Few More Details on the Exoplanet Catalog}
To construct a catalog of exoplanet parameters, we began with the catalog of comprehensive planet data for all confirmed exoplanets maintained by the NASA Exoplanet Archive.
To this, we added the catalog of all candidate Kepler systems, excluding known false positives, downloaded on the same date.
We used the gaia-kepler.fun cross-match catalog created by M. Bedell\footnote{\url{https://gaia-kepler.fun}} to add \textit{Gaia} distance \citep{gaia18} and proper motion data for the \textit{Kepler} candidate hosts.

At the time of writing, the NASA catalogs did not include the samples of planet radii that have most clearly exposed the gap, those of \cite{fulton18} and \cite{eylen18}.
Accurate planetary radii require accurate estimates of stellar radii, which \cite{fulton18} and the \textit{California Kepler Survey} team obtained using spectroscopy for a sample of 1189 stars (1901 planets) and \cite{eylen18} obtained with asteroseismology of 63 host stars (117 planets).
Both of the added catalogs include estimates of stellar mass and the \cite{fulton18} catalog includes planetary instellations and orbital semi-major axes that we used in our analysis.
We added or replaced parameters in the NASA Archive catalogs with values from the \cite{fulton18} and, preferentially, \cite{eylen18} catalogs.

To this catalog, we added data on stellar NUV and FUV emission and corresponding exoplanet NUV and FUV irradiation from archival \textit{GALEX} observations (see Section \ref{sec:uv}).
\textit{GALEX} measurements (not upper limits) of the host star's NUV flux were available for 2384 planets (339 FUV) with radii $<4\ R_\oplus$.
There are many fewer FUV detections because stars are generally much fainter in the \textit{GALEX} FUV than the NUV, making them harder to detect.

\needspace{6em} \section{Vetting the \textit{GALEX} Fluxes}
\label{app:vetting}
\subsection{Anomalies}
For six planets, we estimate an NUV flux received by the planet that is larger than the planet's instellation as given in the NASA archive.
Four of these systems had stellar luminosities given in the archive.
For these, the NUV luminosity we computed was a reasonable fraction of the stellar luminosity ($<$5\%).
Meanwhile, an estimate of the planetary instellation using the stellar luminosity, $L$ and orbital semi-major axis, $a$, (i.e., $L / 4 \pi a^2$) yielded values 20-1000$\times$ the instellation given in the archive.
Hence, we concluded the planetary instellations were in error.
A closer examination of one target, HATS-18b, showed that the appropriate unit conversion had not been applied when the planet parameters were ingested into the NASA archive.
This led us to exclude systems where the catalog instellation deviated by more than a factor of 2 from $L / 4 \pi a^2$ when analyzing $R_p$ versus $S$ or when using detrending according to $S$.

For four stars, the estimated NUV (and, for DP Leo, the estimated FUV) luminosity exceeds 10\% of the catalog bolometric luminosity:
\begin{enumerate}
    \item DP Leo is a cataclysmic variable with an effective temperature of 13,500 K.
    \item Kepler-1611 appears to be a bad \textit{Gaia} match, as the distance given in the NASA archive is 686 pc versus 27,199 pc from the gaia-kepler.fun catalog.
    \item Kepler-953 is also likely a bad \textit{Gaia} match, with a distance of 240 pc in the NASA archive and 1124 pc in the gaia-kepler.fun catalog.
    \item Kepler-416 was likely incorrectly matched to a bright nearby star in the \textit{GALEX} source catalog.
\end{enumerate}
Of the 2813/2964 unique planet hosts with both \textit{Gaia} (from the gaia-kepler.fun catalog) and literature (from the NASA archive) distances, the two differed by more than 50\% in 179 cases (6\%).
In these cases, we favored the NASA archive distance as likely to have been better vetted.

\subsection{Comparison to \cite{shkolnik13}}
\begin{figure}
\includegraphics{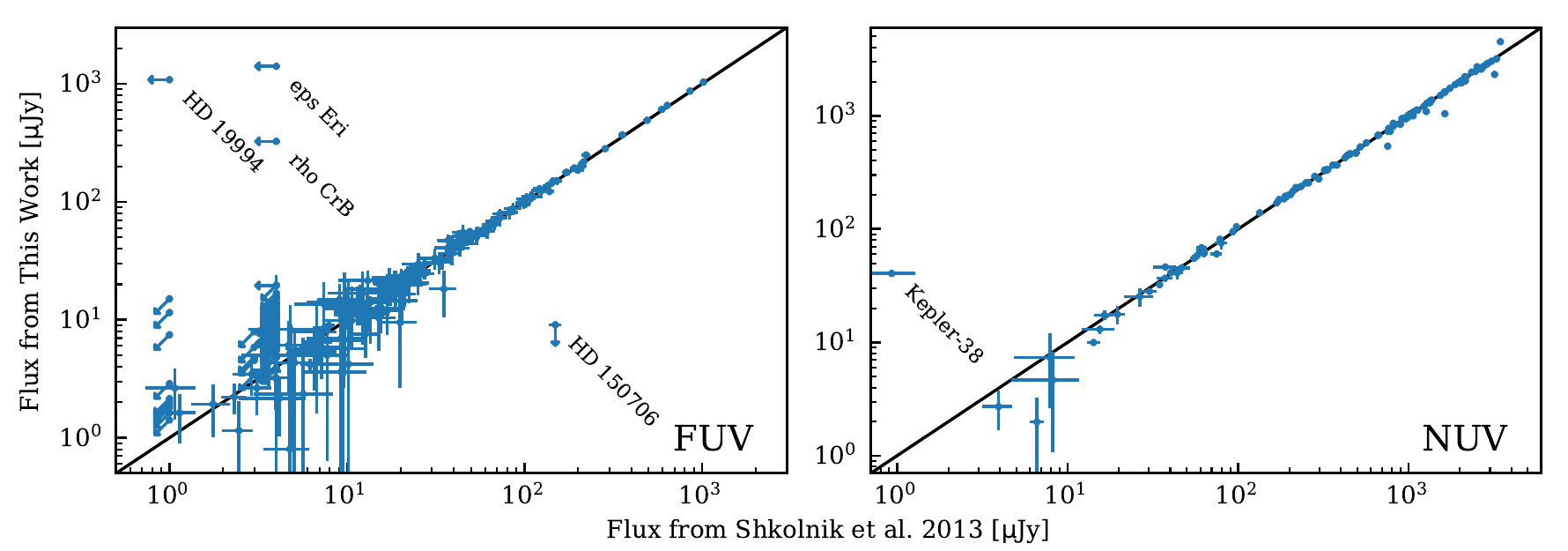}
\caption{Comparison of \textit{GALEX} fluxes retrieved by this work versus \cite{shkolnik13}.
Agreement is good, with most scatter likely due to differences in choice of synthetic aperture, sigma clipping, and method of extracting limits.
In a few cases, the \textit{GALEX} source catalog contained duplicate NUV-only and FUV-only detections of the same source, resulting in false negatives if these duplications were not found.
\label{fig:shkolnik}}
\end{figure}

We compared the fluxes of the objects overlapping with the catalog of \cite{shkolnik13} to identify discrepancies.
Fluxes are compared in Figure \ref{fig:shkolnik}.
The agreement is good, with the small differences likely attributable to the use of different apertures.
\cite{shkolnik13} utilized the ``auto'' aperture values, whereas for this work we used the ``APER\_7'' values.
The auto aperture is a KRON elliptical synthetic aperture of variable size intended to match the source size of potentially resolved galaxies.\footnote{\url{http://www.galex.caltech.edu/wiki/Public:Documentation/Chapter_103\#Guide_to_GALEX_Imaging_Measurements}}
The APER\_7 aperature is a 34.5'' circular synthetic aperture.
We used APER\_7 primarily so that we could apply a saturation correction using the curves published in \cite{morrissey07}, which are provided only for 34.5'' and 3' apertures.

For NUV fluxes, in many cases we estimated upper limits above those estimated by \cite{shkolnik13}.
This is likely due to our use of an upper limit from only a single \textit{GALEX} visit of the target's location.
More restrictive upper limits are possible if all visits are coadded.

Outliers are identified in Figure \ref{fig:shkolnik}.
In the NUV, the only outlier is Kepler-38. 
An estimate of this star's expected NUV flux based on its effective temperature and distance lies between that of this work and \cite{shkolnik13}.
However, querying the \textit{GalexView} online interface,\footnote{http://galex.stsci.edu/galexView/} the nearest object to Kepler-38 has a flux nearer to that of this work.

In the FUV, several outliers exist for which we measured fluxes orders of magnitude above the upper limits provided in \cite{shkolnik13}. 
In all cases, this seems to be a result of the \textit{GALEX} catalog duplicating a source.
This appears in the catalog as two sources within a few arcseconds of each other for which one has an NUV detection without an FUV detection and vice versa.
Although initially these discrepancies with the \cite{shkolnik13} catalog appeared only by happenstance, we afterwards modified our \textit{GALEX} retrieval script to treat these duplicates as a single source.

\subsection{Absolute Magnitude versus Effective Temperature}
\begin{figure}
\includegraphics{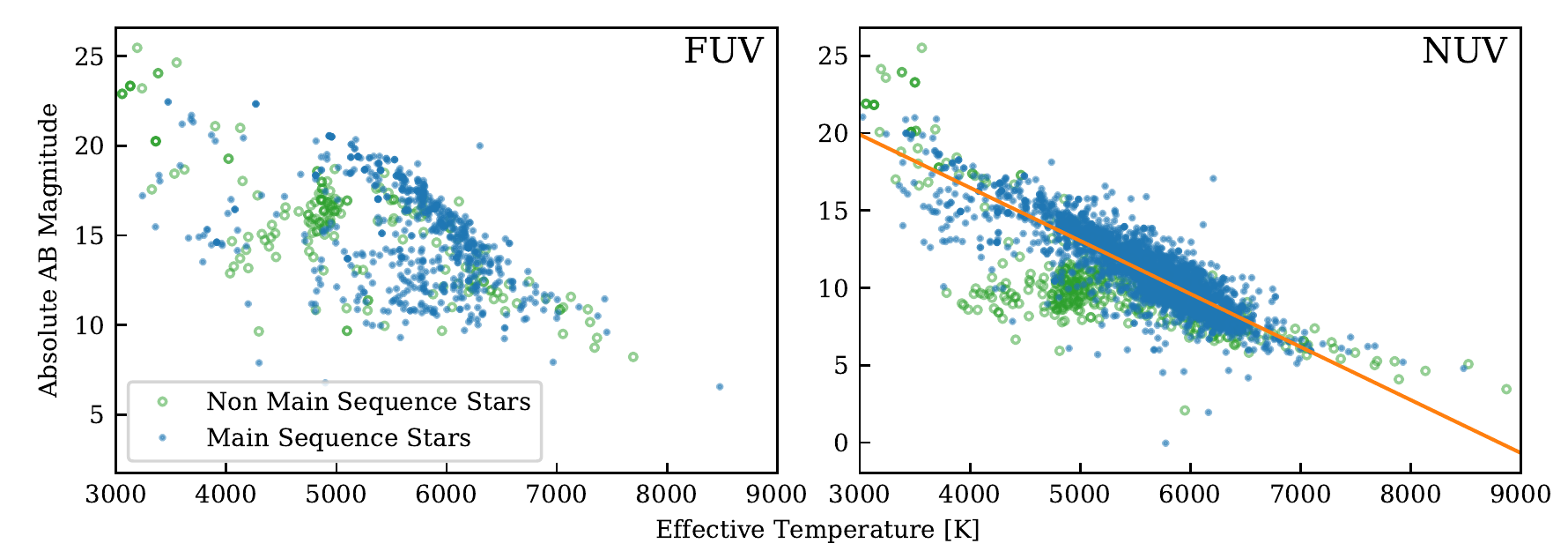}
\caption{Absolute \textit{GALEX} magnitudes of planetary host stars as a function of effective temperature.
\label{fig:UVvsTeff}}
\end{figure}

We also examined absolute \textit{GALEX} magnitude versus effective temperature to search for potential issues with the catalog (Figure \ref{fig:UVvsTeff}).
In general, the sources follow a clear trend.
There is a substantial population of overluminous sources, but a $\log g$ cut shows that most of these sources are not main-sequence stars.
Some of the remainder could be stars caught in a flaring state.
However, the three highest outliers, Kepler-1611, K02962, and K04345, all had measurements from multiple \textit{GALEX} visits and for each these measurements were consistent, suggesting their high fluxes were not simply due to a coincidental flare.
For the NUV measurements, selecting only the stars with $ 4 < \log \left(g\ [\mathrm{cm\ s^{-1}}] \right) < 5 $ and $ 3500\ \mathrm{K} < T_\mathrm{eff} < 7000\ \mathrm{K}$ and fitting a third-order polynomial to the resulting points, we find that 1.7\% are beyond $3\sigma$ from the trend.
The catalog also contains four stars with effective temperatures greater than 10,000~K not shown in Figure \ref{fig:UVvsTeff}, yet \textit{GALEX} magnitudes within the range expected of a main sequence F, G, or K star.
These appear to be all evolved, hot subdwarfs.

\needspace{6em} \section{Tests of Tests: A Proof of Concept Using a Simulated, Idealized Planetary Population}
\label{app:experiments}
Discerning between the evaporation and core-powered mass loss theories using UV data is challenging given the tight correlation between a planet's instellation and its NUV or FUV irradiation.
Hence, we constructed simulated, idealized populations to explore a range of possible tests.
In this section, we describe these experiments using a single population for simplicity.

\begin{figure}
\centering
\includegraphics{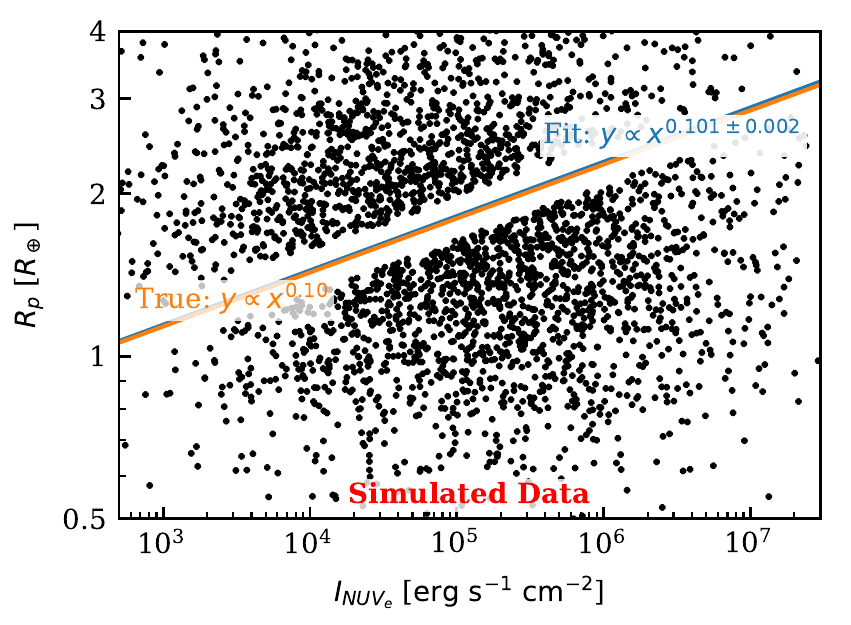}
\caption{A simulated planet population with an injected radius gap used for a numerical proof of concept.
The equation of the injected gap is shown in orange while the retrieved best-fit gap is shown in blue.
The two are nearly indistinguishable for this large ($N = 3000$) sample.
\label{fig:simUV}}
\end{figure}

The simulated population we present consists of a large sample of planets for clarity in this proof of concept, $N=3000$, normally distributed in $R_p$ and \xNUV\ irradiation.
Because our purpose here was merely to explore the functionality of possible population tests, we used an ad hoc method of creating a radius gap that required essentially no time to run rather than using the \cite{owen17} XUV evaporation model.
This ad hoc method consisted of clearing an area of width 0.08 dex in $R_p$ and obeying the relationship $R_{\mathrm{gap}} \propto \IxNUVm^{0.1}$ to mimic the observed relationship with $S$ and \IxNUV\ (Section \ref{sec:results}).
We plot this starting population in Figure \ref{fig:simUV} along with the injected gap and the best fit to that gap produced by our custom algorithm.
Our algorithm recovers the equation of the injected gap well.

\begin{figure}
\centering
\includegraphics{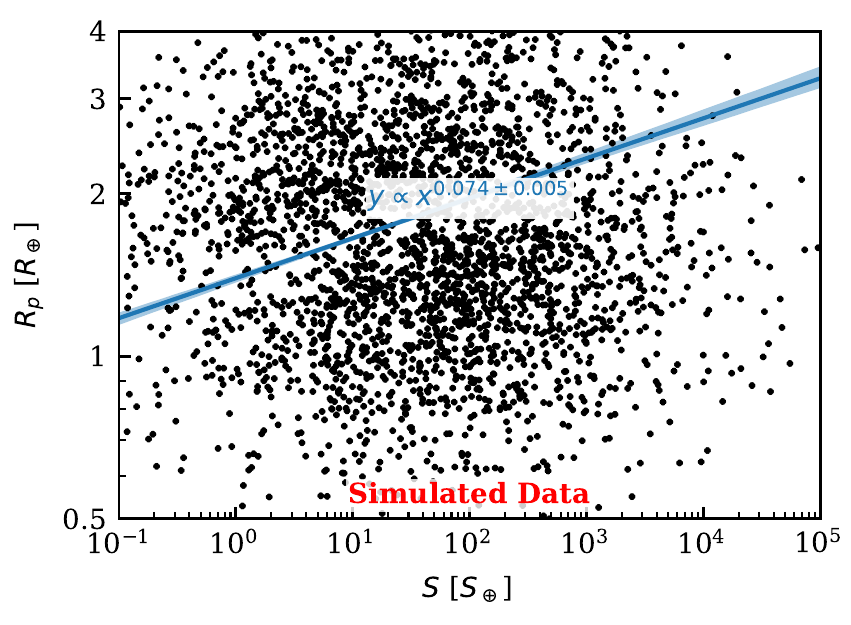}
\caption{The simulated planet population in the $R_p$-$S$ plane with values of $S$ estimated by assuming 0.5 dex scatter in the \lratioN\ of the planetary hosts.
The gap has been partially filled by the scatter.
The slope of the best-fit gap line is lower than in the $R_p$-\IxNUV\ plane because the added scatter causes the $S$ values to span a greater range than the \IxNUV\ values.
\label{fig:simS}}
\end{figure}

Injecting a 1$\sigma$ scatter of 0.5 dex in \lratioN\ (see Introduction) into the simulated planet populations to determine planetary instellations yields the distribution shown in Figure \ref{fig:simS}.
The scatter in \lratioN\ partially fills the gap with planets.
The scatter also stretches the $S$ axis relative to the \IxNUV\ axis, resulting in a weaker dependency of the gap on $S$ (smaller value of the slope of the gap).

\begin{figure}
\includegraphics{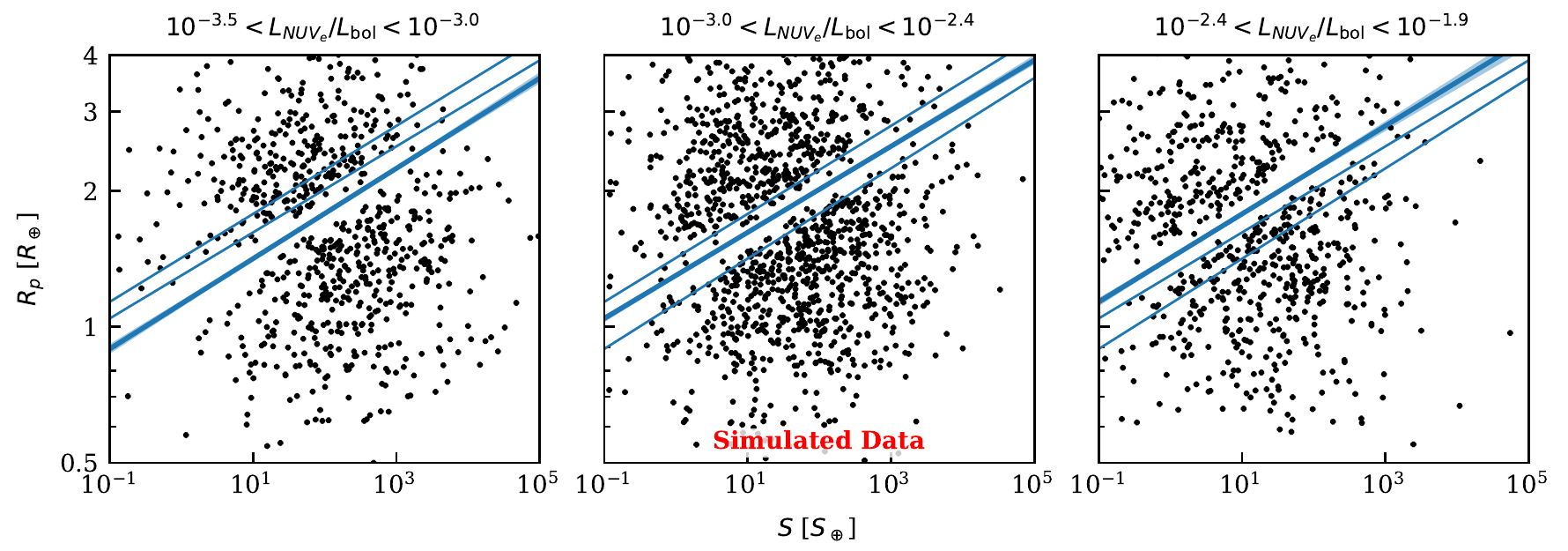}
\caption{The simulated planet population, split into three subpopulations based on even ranges of \lratioN\, and plotted in the $R_p$-$S$ plane.
The location of the gap increases in $R_p$ for the subpopulations with larger \lratioN, as shown by the best-fit lines.
The fits to all three subpopulations are shown in each panel, with a thicker line with error region indicating the fit to the plotted population.
\label{fig:simSplits}}
\end{figure}

This is the theoretical starting point at which we begin to try to uncover the fact that the gap is better described by the \xNUV\ irradiation of the planets than their instellation.
We begin by describing the test that is, perhaps, the easiest to interpret.
For this test, we binned the planets into three separate populations based on their \lratioN.
Then, we fit the gap in the $R_p$-$S$ plane for each of these populations.
Because it was the \xNUV\ radiation that created this fictional gap, the gap should move to higher $R_p$ for populations with higher average \lratioN.
Indeed, this was the result, as shown in Figure \ref{fig:simSplits}.

\begin{figure}
\centering
\includegraphics{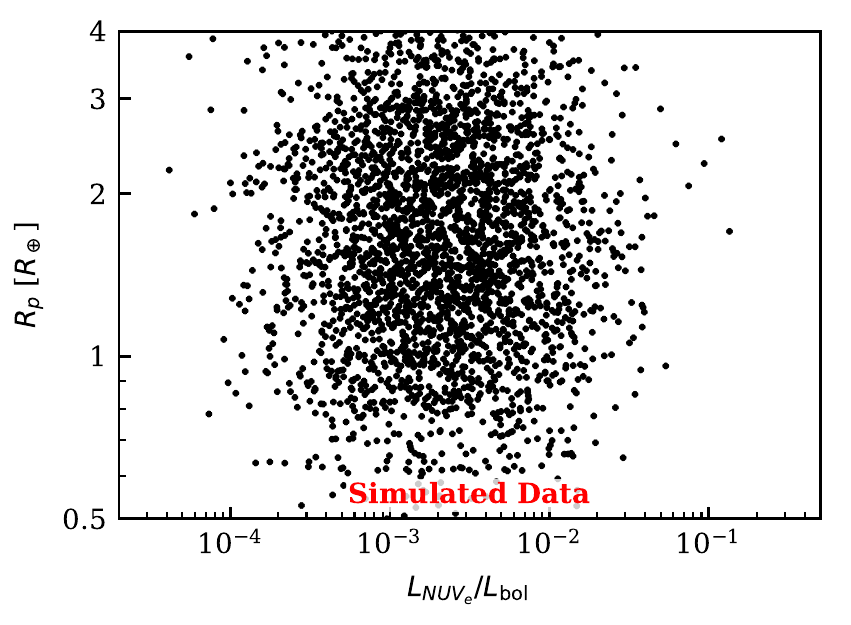}
\caption{The simulated planet population plotted in the $R_p$-\lratioN\ plane.
The gap has been obscured due to the scatter from the large range of possible \xNUV\ irradiation for planets with similar \lratioN, which drives the radius of the gap in this idealized sample.
\label{fig:simxNUV_bol}}
\end{figure}

Next we attempted to discern a gap in the entire population with a dependency on \lratioN.
However, the large range in possible \IxNUV\ values for planets with similar \lratioN, associated with a large range of gap radii, acts to obscure the gap in the $R_p$-\lratioN\ plane (Figure \ref{fig:simxNUV_bol}).

\begin{figure}
\centering
\includegraphics{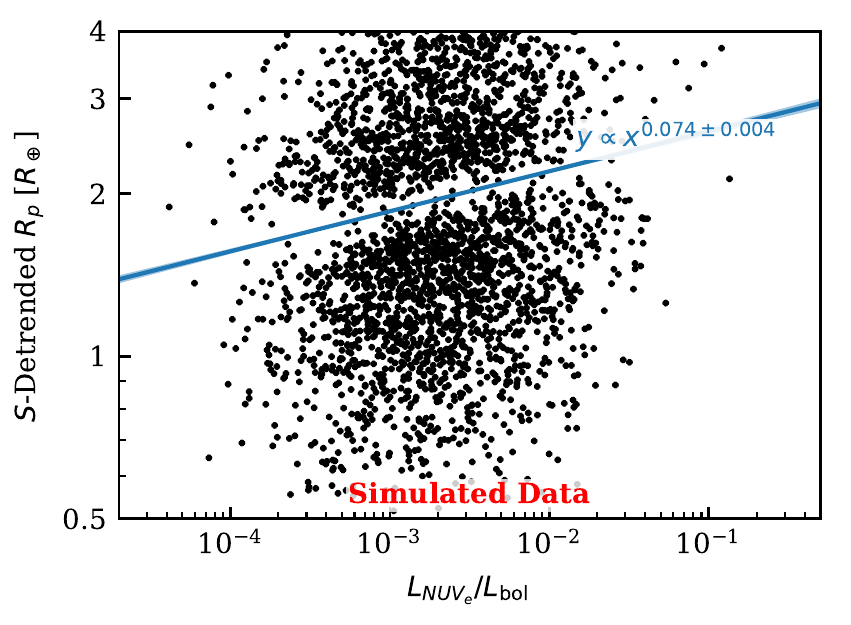}
\caption{The simulated planet population plotted in the shifted $R_p$-\lratioN\ plane, where ``detrended'' indicates that the $R_p$ values have been adjusted according to the fit to the gap in the $R_p$-$S$ plane.
This removes the dependency of the radius gap on the instellation of planets and allows the residual dependency on the fraction of that irradiation that is better correlated with atmospheric stripping to show through.
\label{fig:simBest}}
\end{figure}

The obscuration of the gap in the $R_p$-\lratioN\ plane can be effectively removed by first detrending the planetary radii based on the $R_p$-$S$ dependency of the gap.
By detrending, we mean subtracting the log planetary radii given by the  $R_p$-$S$ fit from the actual log planetary radii.
To avoid confusing the casual reader, we then add back the log radius at the reference x-value of the fit, e.g., 1.8~$R_p$ at 100~$S_\oplus$, in log space.
This accounts for the effect of the different instellation of a planet, based on its proximity to its host star and its host star's luminosity, on the maximum size of the planets whose atmospheres can be stripped.
All that is left after this detrending is the effect from the fraction of the light irradiating the planet that can power atmospheric stripping, or, in this case, a proxy for that fraction: \lratioN.
Larger fractions of atmosphere-stripping radiation push the gap to larger planetary radii, and this relationship shows through after $S$ detrending (Figure \ref{fig:simBest}).

\begin{figure*}
\includegraphics{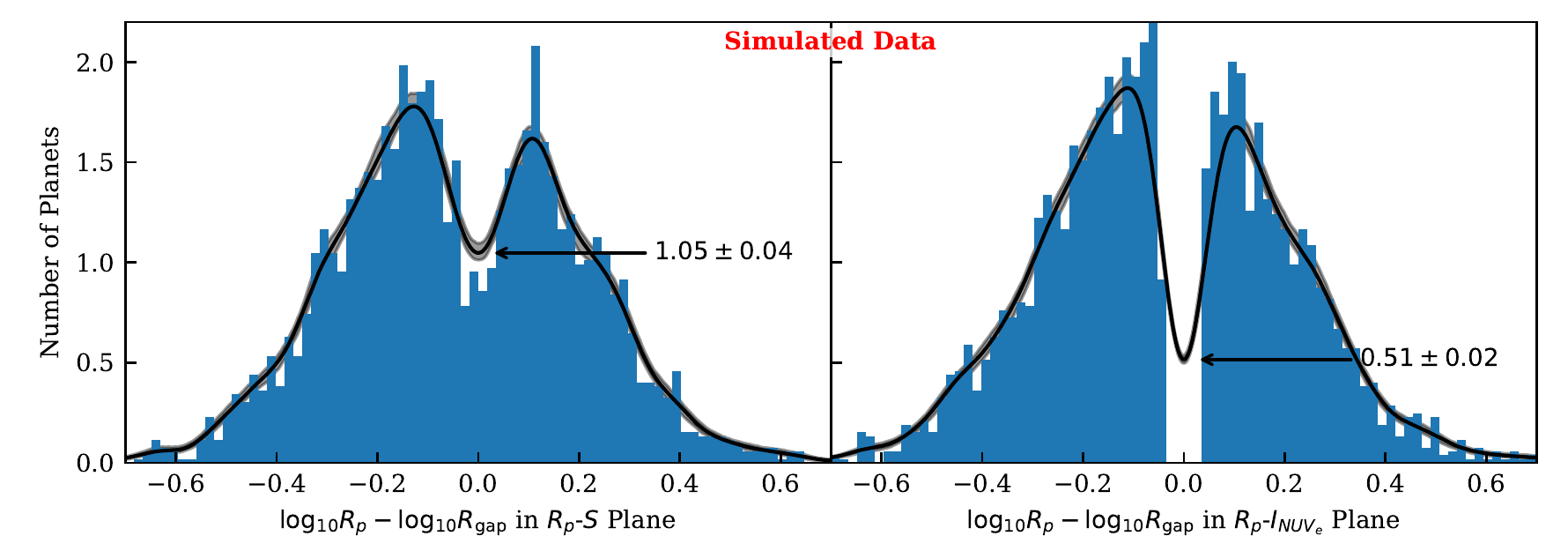}
\caption{A comparison of the depth of the gap in the $R_p$-$S$ and the $R_p$-\IxNUV\ planes.
The x-axes are the values of the planetary radii after subtraction of the best-fit radius gap, corresponding to Figures \ref{fig:simUV} and \ref{fig:simS}.
The black line is a kernel density estimate of the occurrence frequency and the gray area is the 16th to 84th percentile region from estimates for bootstrapped samples.
The depth of the gap is significantly greater in the $R_p$-\IxNUV\ plane, but we caution that the application of this test to true planetary samples will likely be less powerful because the edges of the true gap will not be step functions.
\label{fig:simDepth}}
\end{figure*}

Finally, another straightforward test is to compare how clean the gap is in the $R_p$-$S$ and $R_p$-\IxNUV\ planes.
This could be quantified in a number of ways, but the simplest is to measure the depth of the gap along the best fit line.
More precisely, this utilizes gap-subtracted planetary radii, where the log of the best-fit radius gap has been subtracted from the log of the planetary radii.
However, the utility of this test is highly dependent on the assumed shape of the gap.
In this simulation, the gap is highly idealized as a step function.
In reality, the gap, even when plotted as a function of some causal variable, is likely to be partially filled (e.g., \citealt{owen17}).
This could make changes in the gap depth between a causal variable and some correlate of that variable hard to discern.
Nonetheless, the test was worth performing, and, in this idealized case, would have revealed a clear difference between the depth of the gap in the $R_p$-$S$ and $R_p$-\IxNUV\ plane that favors \IxNUV\ as better characterizing the gap (Figure \ref{fig:simDepth}).

\needspace{6em} \section{Individual Gap Fit Plots for Planet Populations Binned by \lratioN\ and \M}
\label{app:splitfits}
\begin{figure*}
\includegraphics{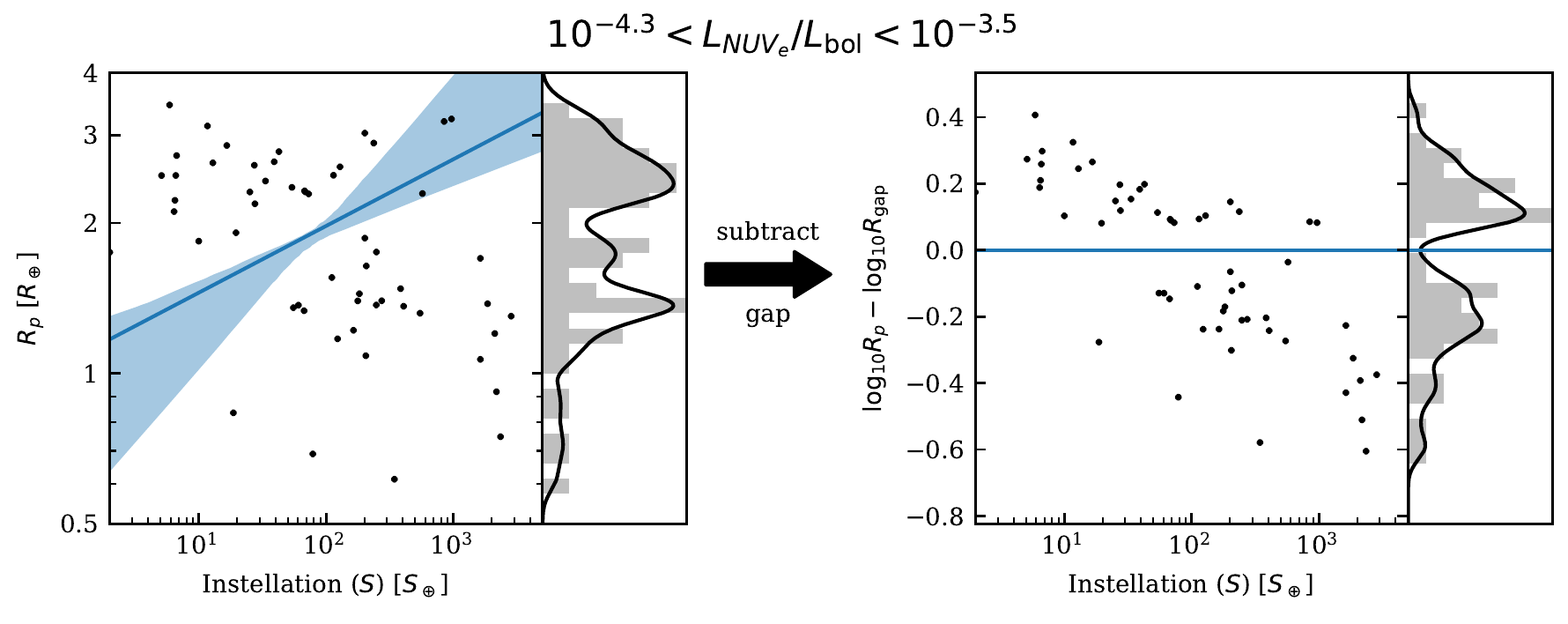}
\caption{Planets in the $R_p$-$S$ plane for the population of planets orbiting hosts in the \textbf{lower} third of the \lratioN\ distribution.
\fitCaptionBoiler
\label{fig:SlowUV}}
\end{figure*}

\begin{figure*}
\includegraphics{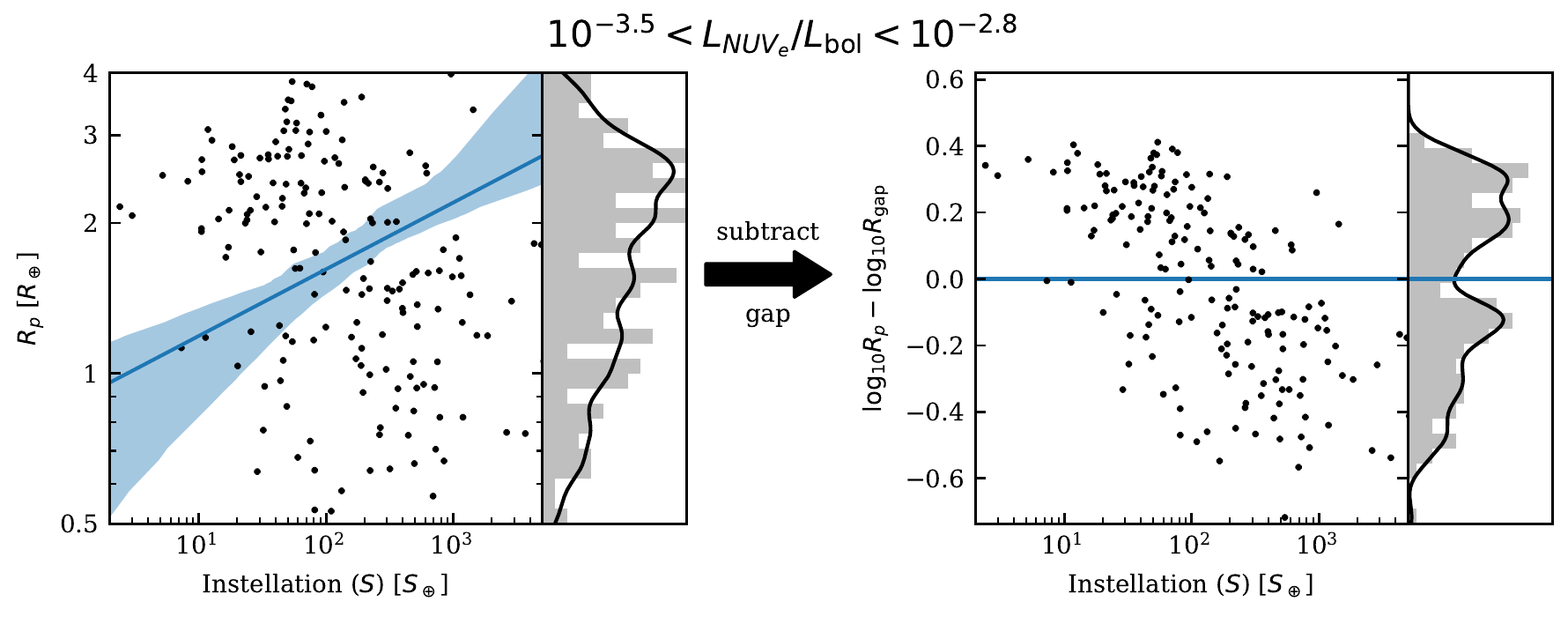}
\caption{Planets in the $R_p$-$S$ plane for the population of planets orbiting hosts in the \textbf{middle} third of the \lratioN\ distribution.
\fitCaptionBoiler
\label{fig:SmidUV}}
\end{figure*}

\begin{figure*}
\includegraphics{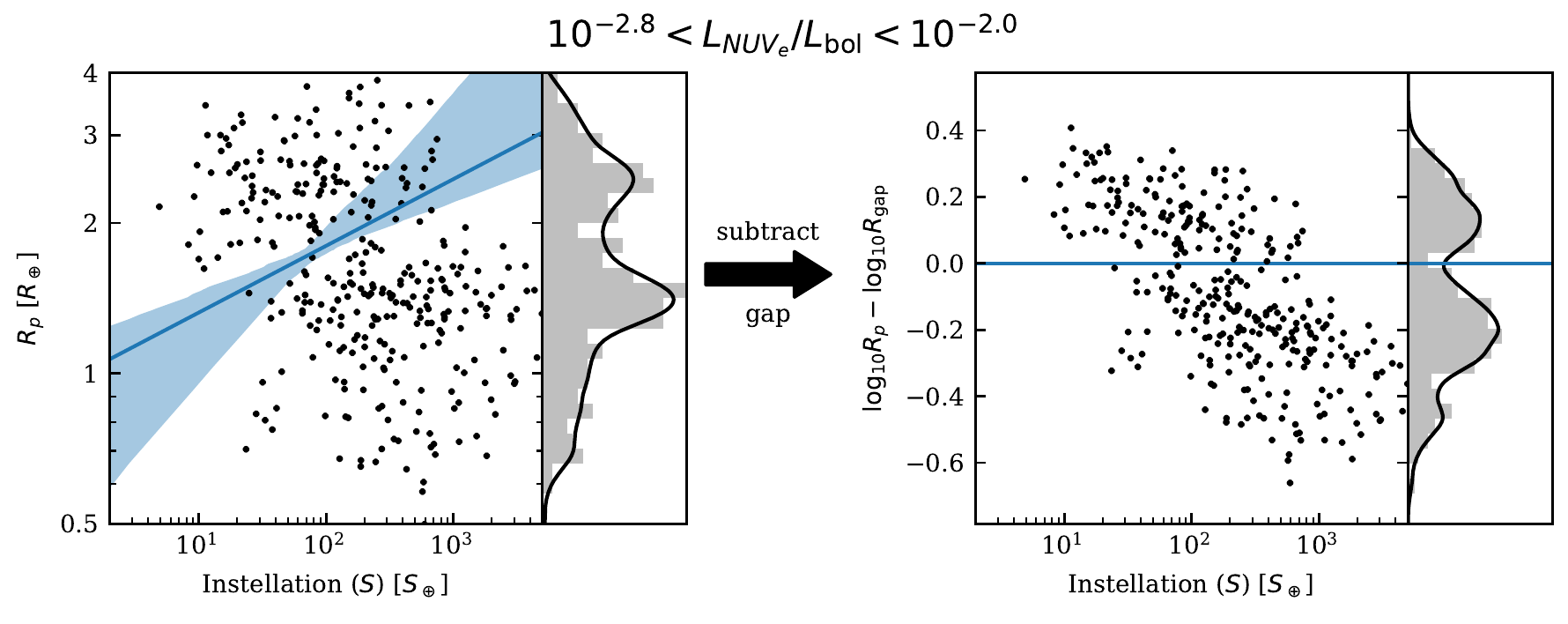}
\caption{Planets in the $R_p$-$S$ plane for the population of planets orbiting hosts in the \textbf{upper} third of the \lratioN\ distribution.
\fitCaptionBoiler
\label{fig:ShiUV}}
\end{figure*}

\begin{figure*}
\includegraphics{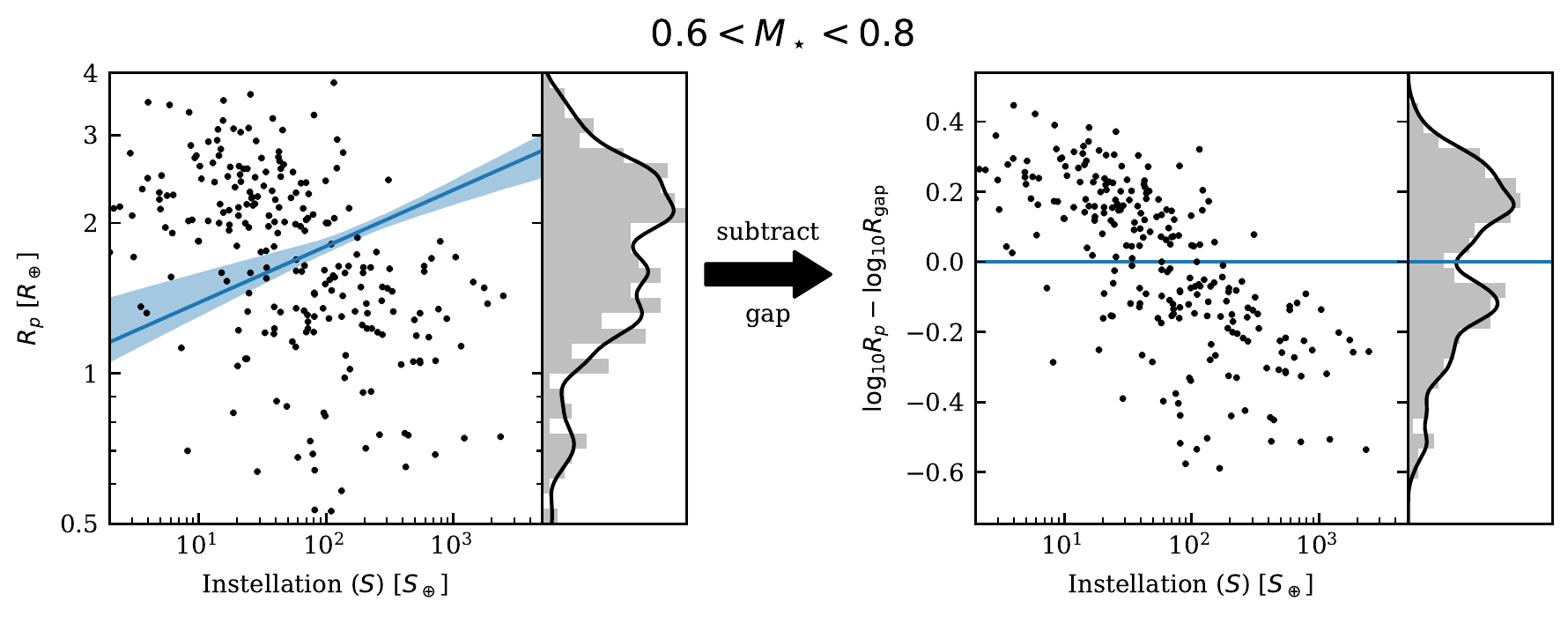}
\caption{Planets in the $R_p$-$S$ plane for the population of planets orbiting hosts in the \textbf{lower} third of the host star mass distribution.
\fitCaptionBoiler
\label{fig:SlowMass}}
\end{figure*}

\begin{figure*}
\includegraphics{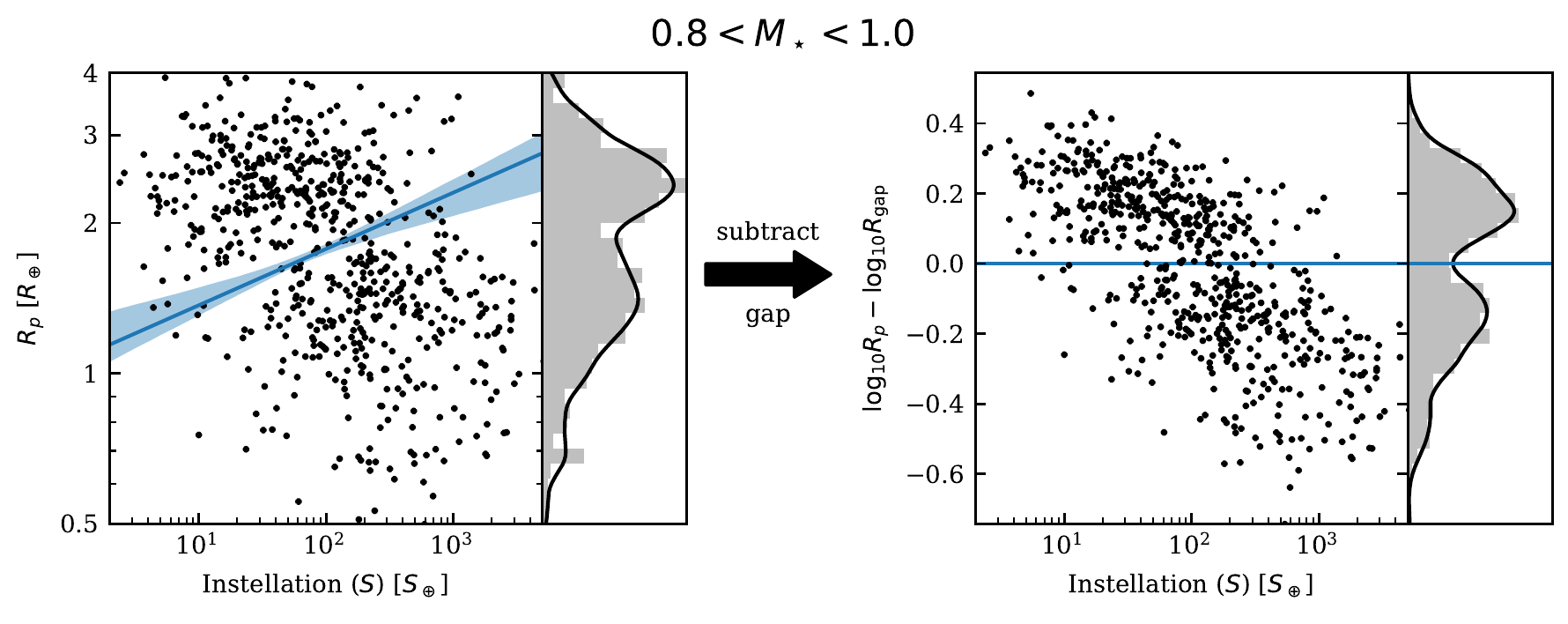}
\caption{Planets in the $R_p$-$S$ plane for the population of planets orbiting hosts in the \textbf{middle} third of the host star mass distribution.
\fitCaptionBoiler
\label{fig:SmidMass}}
\end{figure*}

\begin{figure*}
\includegraphics{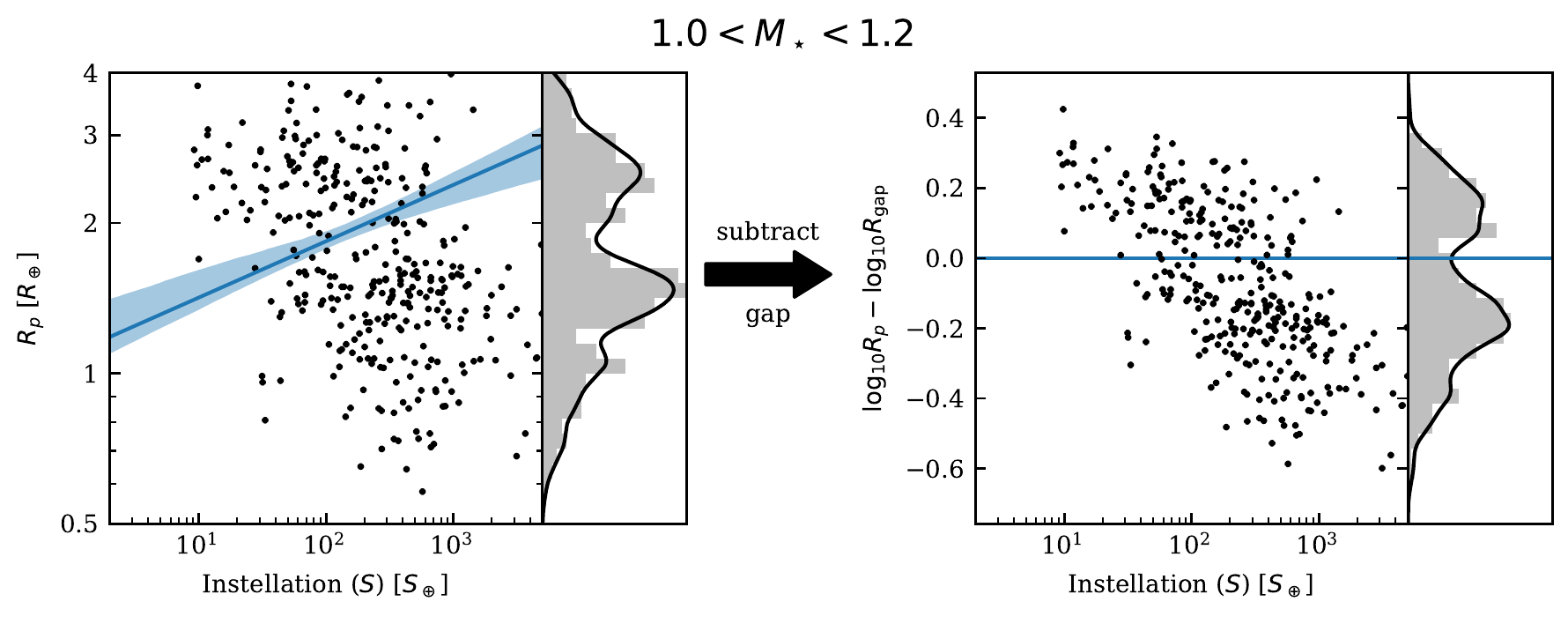}
\caption{Planets in the $R_p$-$S$ plane for the population of planets orbiting hosts in the \textbf{upper} third of the host star mass distribution.
\fitCaptionBoiler
\label{fig:ShiMass}}
\end{figure*}

In Section \ref{sec:results} we showed in Figures \ref{fig:UVsplitfits} and  \ref{fig:Msplitfits} three fits to the radius gap in the $R_p$-$S$ plane for three populations of planets sorted by their \lratioN\ and host star mass.
We show the individual fits to the radius gap for these populations in Figures \ref{fig:SlowUV} - \ref{fig:ShiMass}.

\bibliography{refs}{}

\begin{thebibliography}{}
\expandafter\ifx\csname natexlab\endcsname\relax\def\natexlab#1{#1}\fi
\providecommand{\url}[1]{\href{#1}{#1}}

\bibitem[{{Astropy Collaboration} {et~al.}(2013){Astropy Collaboration},
  {Robitaille}, {Tollerud}, {Greenfield}, {Droettboom}, {Bray}, {Aldcroft},
  {Davis}, {Ginsburg}, {Price-Whelan}, {Kerzendorf}, {Conley}, {Crighton},
  {Barbary}, {Muna}, {Ferguson}, {Grollier}, {Parikh}, {Nair}, {Unther},
  {Deil}, {Woillez}, {Conseil}, {Kramer}, {Turner}, {Singer}, {Fox}, {Weaver},
  {Zabalza}, {Edwards}, {Azalee Bostroem}, {Burke}, {Casey}, {Crawford},
  {Dencheva}, {Ely}, {Jenness}, {Labrie}, {Lim}, {Pierfederici}, {Pontzen},
  {Ptak}, {Refsdal}, {Servillat}, \& {Streicher}}]{astropy13}
{Astropy Collaboration}, {Robitaille}, T.~P., {Tollerud}, E.~J., {et~al.} 2013,
  \aap, 558, A33

\bibitem[{{Epstein} \& {Pinsonneault}(2014)}]{epstein14}
{Epstein}, C.~R., \& {Pinsonneault}, M.~H. 2014, \apj, 780, 159

\bibitem[{{France} {et~al.}(2018){France}, {Arulanantham}, {Fossati}, {Lanza},
  {Loyd}, {Redfield}, \& {Schneider}}]{france18}
{France}, K., {Arulanantham}, N., {Fossati}, L., {et~al.} 2018, \apjs, 239, 16

\bibitem[{{Fulton} \& {Petigura}(2018)}]{fulton18}
{Fulton}, B.~J., \& {Petigura}, E.~A. 2018, \aj, 156, 264

\bibitem[{{Fulton} {et~al.}(2017){Fulton}, {Petigura}, {Howard}, {Isaacson},
  {Marcy}, {Cargile}, {Hebb}, {Weiss}, {Johnson}, {Morton}, {Sinukoff},
  {Crossfield}, \& {Hirsch}}]{fulton17}
{Fulton}, B.~J., {Petigura}, E.~A., {Howard}, A.~W., {et~al.} 2017, \aj, 154,
  109

\bibitem[{{Gaia Collaboration} {et~al.}(2018){Gaia Collaboration}, {Brown},
  {Vallenari}, {Prusti}, {de Bruijne}, {Babusiaux}, {Bailer-Jones}, {Biermann},
  {Evans}, {Eyer}, {Jansen}, {Jordi}, {Klioner}, {Lammers}, {Lindegren},
  {Luri}, {Mignard}, {Panem}, {Pourbaix}, {Randich}, {Sartoretti}, {Siddiqui},
  {Soubiran}, {van Leeuwen}, {Walton}, {Arenou}, {Bastian}, {Cropper},
  {Drimmel}, {Katz}, {Lattanzi}, {Bakker}, {Cacciari}, {Casta{\~n}eda},
  {Chaoul}, {Cheek}, {De Angeli}, {Fabricius}, {Guerra}, {Holl}, {Masana},
  {Messineo}, {Mowlavi}, {Nienartowicz}, {Panuzzo}, {Portell}, {Riello},
  {Seabroke}, {Tanga}, {Th{\'e}venin}, {Gracia-Abril}, {Comoretto},
  {Garcia-Reinaldos}, {Teyssier}, {Altmann}, {Andrae}, {Audard},
  {Bellas-Velidis}, {Benson}, {Berthier}, {Blomme}, {Burgess}, {Busso},
  {Carry}, {Cellino}, {Clementini}, {Clotet}, {Creevey}, {Davidson}, {De
  Ridder}, {Delchambre}, {Dell'Oro}, {Ducourant},
  {Fern{\'a}ndez-Hern{\'a}ndez}, {Fouesneau}, {Fr{\'e}mat}, {Galluccio},
  {Garc{\'\i}a-Torres}, {Gonz{\'a}lez-N{\'u}{\~n}ez}, {Gonz{\'a}lez-Vidal},
  {Gosset}, {Guy}, {Halbwachs}, {Hambly}, {Harrison}, {Hern{\'a}ndez},
  {Hestroffer}, {Hodgkin}, {Hutton}, {Jasniewicz}, {Jean-Antoine-Piccolo},
  {Jordan}, {Korn}, {Krone-Martins}, {Lanzafame}, {Lebzelter}, {L{\"o}ffler},
  {Manteiga}, {Marrese}, {Mart{\'\i}n-Fleitas}, {Moitinho}, {Mora}, {Muinonen},
  {Osinde}, {Pancino}, {Pauwels}, {Petit}, {Recio-Blanco}, {Richards},
  {Rimoldini}, {Robin}, {Sarro}, {Siopis}, {Smith}, {Sozzetti}, {S{\"u}veges},
  {Torra}, {van Reeven}, {Abbas}, {Abreu Aramburu}, {Accart}, {Aerts},
  {Altavilla}, {{\'A}lvarez}, {Alvarez}, {Alves}, {Anderson}, {Andrei},
  {Anglada Varela}, {Antiche}, {Antoja}, {Arcay}, {Astraatmadja}, {Bach},
  {Baker}, {Balaguer-N{\'u}{\~n}ez}, {Balm}, {Barache}, {Barata}, {Barbato},
  {Barblan}, {Barklem}, {Barrado}, {Barros}, {Barstow}, {Bartholom{\'e}
  Mu{\~n}oz}, {Bassilana}, {Becciani}, {Bellazzini}, {Berihuete}, {Bertone},
  {Bianchi}, {Bienaym{\'e}}, {Blanco-Cuaresma}, {Boch}, {Boeche}, {Bombrun},
  {Borrachero}, {Bossini}, {Bouquillon}, {Bourda}, {Bragaglia}, {Bramante},
  {Breddels}, {Bressan}, {Brouillet}, {Br{\"u}semeister}, {Brugaletta},
  {Bucciarelli}, {Burlacu}, {Busonero}, {Butkevich}, {Buzzi}, {Caffau},
  {Cancelliere}, {Cannizzaro}, {Cantat-Gaudin}, {Carballo}, {Carlucci},
  {Carrasco}, {Casamiquela}, {Castellani}, {Castro-Ginard}, {Charlot},
  {Chemin}, {Chiavassa}, {Cocozza}, {Costigan}, {Cowell}, {Crifo}, {Crosta},
  {Crowley}, {Cuypers}, {Dafonte}, {Damerdji}, {Dapergolas}, {David}, {David},
  {de Laverny}, {De Luise}, {De March}, {de Martino}, {de Souza}, {de Torres},
  {Debosscher}, {del Pozo}, {Delbo}, {Delgado}, {Delgado}, {Di Matteo},
  {Diakite}, {Diener}, {Distefano}, {Dolding}, {Drazinos}, {Dur{\'a}n},
  {Edvardsson}, {Enke}, {Eriksson}, {Esquej}, {Eynard Bontemps}, {Fabre},
  {Fabrizio}, {Faigler}, {Falc{\~a}o}, {Farr{\`a}s Casas}, {Federici},
  {Fedorets}, {Fernique}, {Figueras}, {Filippi}, {Findeisen}, {Fonti},
  {Fraile}, {Fraser}, {Fr{\'e}zouls}, {Gai}, {Galleti}, {Garabato},
  {Garc{\'\i}a-Sedano}, {Garofalo}, {Garralda}, {Gavel}, {Gavras}, {Gerssen},
  {Geyer}, {Giacobbe}, {Gilmore}, {Girona}, {Giuffrida}, {Glass}, {Gomes},
  {Granvik}, {Gueguen}, {Guerrier}, {Guiraud}, {Guti{\'e}rrez-S{\'a}nchez},
  {Haigron}, {Hatzidimitriou}, {Hauser}, {Haywood}, {Heiter}, {Helmi}, {Heu},
  {Hilger}, {Hobbs}, {Hofmann}, {Holland}, {Huckle}, {Hypki}, {Icardi},
  {Jan{\ss}en}, {Jevardat de Fombelle}, {Jonker}, {Juh{\'a}sz}, {Julbe},
  {Karampelas}, {Kewley}, {Klar}, {Kochoska}, {Kohley}, {Kolenberg},
  {Kontizas}, {Kontizas}, {Koposov}, {Kordopatis}, {Kostrzewa-Rutkowska},
  {Koubsky}, {Lambert}, {Lanza}, {Lasne}, {Lavigne}, {Le Fustec}, {Le
  Poncin-Lafitte}, {Lebreton}, {Leccia}, {Leclerc}, {Lecoeur-Taibi},
  {Lenhardt}, {Leroux}, {Liao}, {Licata}, {Lindstr{\o}m}, {Lister}, {Livanou},
  {Lobel}, {L{\'o}pez}, {Managau}, {Mann}, {Mantelet}, {Marchal}, {Marchant},
  {Marconi}, {Marinoni}, {Marschalk{\'o}}, {Marshall}, {Martino}, {Marton},
  {Mary}, {Massari}, {Matijevi{\v{c}}}, {Mazeh}, {McMillan}, {Messina},
  {Michalik}, {Millar}, {Molina}, {Molinaro}, {Moln{\'a}r}, {Montegriffo},
  {Mor}, {Morbidelli}, {Morel}, {Morris}, {Mulone}, {Muraveva}, {Musella},
  {Nelemans}, {Nicastro}, {Noval}, {O'Mullane}, {Ord{\'e}novic},
  {Ord{\'o}{\~n}ez-Blanco}, {Osborne}, {Pagani}, {Pagano}, {Pailler},
  {Palacin}, {Palaversa}, {Panahi}, {Pawlak}, {Piersimoni}, {Pineau}, {Plachy},
  {Plum}, {Poggio}, {Poujoulet}, {Pr{\v{s}}a}, {Pulone}, {Racero}, {Ragaini},
  {Rambaux}, {Ramos-Lerate}, {Regibo}, {Reyl{\'e}}, {Riclet}, {Ripepi}, {Riva},
  {Rivard}, {Rixon}, {Roegiers}, {Roelens}, {Romero-G{\'o}mez}, {Rowell},
  {Royer}, {Ruiz-Dern}, {Sadowski}, {Sagrist{\`a} Sell{\'e}s}, {Sahlmann},
  {Salgado}, {Salguero}, {Sanna}, {Santana-Ros}, {Sarasso}, {Savietto},
  {Schultheis}, {Sciacca}, {Segol}, {Segovia}, {S{\'e}gransan}, {Shih},
  {Siltala}, {Silva}, {Smart}, {Smith}, {Solano}, {Solitro}, {Sordo}, {Soria
  Nieto}, {Souchay}, {Spagna}, {Spoto}, {Stampa}, {Steele},
  {Steidelm{\"u}ller}, {Stephenson}, {Stoev}, {Suess}, {Surdej}, {Szabados},
  {Szegedi-Elek}, {Tapiador}, {Taris}, {Tauran}, {Taylor}, {Teixeira},
  {Terrett}, {Teyssand ier}, {Thuillot}, {Titarenko}, {Torra Clotet}, {Turon},
  {Ulla}, {Utrilla}, {Uzzi}, {Vaillant}, {Valentini}, {Valette}, {van Elteren},
  {Van Hemelryck}, {van Leeuwen}, {Vaschetto}, {Vecchiato}, {Veljanoski},
  {Viala}, {Vicente}, {Vogt}, {von Essen}, {Voss}, {Votruba}, {Voutsinas},
  {Walmsley}, {Weiler}, {Wertz}, {Wevers}, {Wyrzykowski}, {Yoldas},
  {{\v{Z}}erjal}, {Ziaeepour}, {Zorec}, {Zschocke}, {Zucker}, {Zurbach}, \&
  {Zwitter}}]{gaia18}
{Gaia Collaboration}, {Brown}, A.~G.~A., {Vallenari}, A., {et~al.} 2018, \aap,
  616, A1

\bibitem[{{Ginzburg} {et~al.}(2016){Ginzburg}, {Schlichting}, \&
  {Sari}}]{ginzburg16}
{Ginzburg}, S., {Schlichting}, H.~E., \& {Sari}, R. 2016, \apj, 825, 29

\bibitem[{{Ginzburg} {et~al.}(2018){Ginzburg}, {Schlichting}, \&
  {Sari}}]{ginzburg18}
---. 2018, \mnras, 476, 759

\bibitem[{{Gupta} \& {Schlichting}(2019{\natexlab{a}})}]{gupta19a}
{Gupta}, A., \& {Schlichting}, H.~E. 2019{\natexlab{a}}, \mnras, 487, 24

\bibitem[{{Gupta} \& {Schlichting}(2019{\natexlab{b}})}]{gupta19b}
---. 2019{\natexlab{b}}, arXiv e-prints, arXiv:1907.03732

\bibitem[{{Hinkel} {et~al.}(2017){Hinkel}, {Mamajek}, {Turnbull}, {Osby},
  {Shkolnik}, {Smith}, {Klimasewski}, {Somers}, \& {Desch}}]{hinkel17}
{Hinkel}, N.~R., {Mamajek}, E.~E., {Turnbull}, M.~C., {et~al.} 2017, \apj, 848,
  34

\bibitem[{{Hunten}(1993)}]{hunten93}
{Hunten}, D.~M. 1993, Science, 259, 915

\bibitem[{{Jackson} {et~al.}(2012){Jackson}, {Davis}, \&
  {Wheatley}}]{jackson12}
{Jackson}, A.~P., {Davis}, T.~A., \& {Wheatley}, P.~J. 2012, \mnras, 422, 2024

\bibitem[{{Kubyshkina} {et~al.}(2019){Kubyshkina}, {Cubillos}, {Fossati},
  {Erkaev}, {Johnstone}, {Kislyakova}, {Lammer}, {Lendl}, {Odert}, \&
  {G{\"u}del}}]{kubyshkina19}
{Kubyshkina}, D., {Cubillos}, P.~E., {Fossati}, L., {et~al.} 2019, \apj, 879,
  26

\bibitem[{{Lopez} \& {Fortney}(2013)}]{lopez13}
{Lopez}, E.~D., \& {Fortney}, J.~J. 2013, \apj, 776, 2

\bibitem[{{Lopez} \& {Fortney}(2014)}]{lopez14}
---. 2014, \apj, 792, 1

\bibitem[{{Lopez} {et~al.}(2012){Lopez}, {Fortney}, \& {Miller}}]{lopez12}
{Lopez}, E.~D., {Fortney}, J.~J., \& {Miller}, N. 2012, \apj, 761, 59

\bibitem[{{Luger} {et~al.}(2015){Luger}, {Barnes}, {Lopez}, {Fortney},
  {Jackson}, \& {Meadows}}]{luger15}
{Luger}, R., {Barnes}, R., {Lopez}, E., {et~al.} 2015, Astrobiology, 15, 57

\bibitem[{{MacDonald}(2019)}]{macdonald19}
{MacDonald}, M.~G. 2019, \mnras, 1417

\bibitem[{{Martinez} {et~al.}(2019){Martinez}, {Cunha}, {Ghezzi}, \&
  {Smith}}]{martinez19}
{Martinez}, C.~F., {Cunha}, K., {Ghezzi}, L., \& {Smith}, V.~V. 2019, \apj,
  875, 29

\bibitem[{{McDonald} {et~al.}(2019){McDonald}, {Kreidberg}, \&
  {Lopez}}]{mcdonald19}
{McDonald}, G.~D., {Kreidberg}, L., \& {Lopez}, E. 2019, \apj, 876, 22

\bibitem[{{Morrissey} {et~al.}(2007){Morrissey}, {Conrow}, {Barlow}, {Small},
  {Seibert}, {Wyder}, {Budav{\'a}ri}, {Arnouts}, {Friedman}, {Forster},
  {Martin}, {Neff}, {Schiminovich}, {Bianchi}, {Donas}, {Heckman}, {Lee},
  {Madore}, {Milliard}, {Rich}, {Szalay}, {Welsh}, \& {Yi}}]{morrissey07}
{Morrissey}, P., {Conrow}, T., {Barlow}, T.~A., {et~al.} 2007, \apjs, 173, 682

\bibitem[{{Owen} \& {Campos Estrada}(2019 in press)}]{owen19x}
{Owen}, J.~E., \& {Campos Estrada}, B. 2019 in press, \mnras

\bibitem[{{Owen} \& {Jackson}(2012)}]{owen12}
{Owen}, J.~E., \& {Jackson}, A.~P. 2012, \mnras, 425, 2931

\bibitem[{{Owen} \& {Wu}(2013)}]{owen13}
{Owen}, J.~E., \& {Wu}, Y. 2013, \apj, 775, 105

\bibitem[{{Owen} \& {Wu}(2017)}]{owen17}
---. 2017, \apj, 847, 29

\bibitem[{{Richey-Yowell} {et~al.}(2019){Richey-Yowell}, {Shkolnik},
  {Schneider}, {Osby}, {Barman}, \& {Meadows}}]{richey19}
{Richey-Yowell}, T., {Shkolnik}, E.~L., {Schneider}, A.~C., {et~al.} 2019,
  \apj, 872, 17

\bibitem[{{Schneider} \& {Shkolnik}(2018)}]{schneider18}
{Schneider}, A.~C., \& {Shkolnik}, E.~L. 2018, \aj, 155, 122

\bibitem[{{Shkolnik}(2013)}]{shkolnik13}
{Shkolnik}, E.~L. 2013, \apj, 766, 9

\bibitem[{{Shkolnik} \& {Barman}(2014)}]{shkolnik14}
{Shkolnik}, E.~L., \& {Barman}, T.~S. 2014, \aj, 148, 64

\bibitem[{{Swain} {et~al.}(2018){Swain}, {Estrela}, {Sotin}, {Roudier}, \&
  {Zellem}}]{swain18}
{Swain}, M., {Estrela}, R., {Sotin}, C., {Roudier}, G., \& {Zellem}, R. 2018,
  arXiv e-prints, arXiv:1811.07919

\bibitem[{{Tu} {et~al.}(2015){Tu}, {Johnstone}, {G{\"u}del}, \&
  {Lammer}}]{tu15}
{Tu}, L., {Johnstone}, C.~P., {G{\"u}del}, M., \& {Lammer}, H. 2015, \aap, 577,
  L3

\bibitem[{{Van Eylen} {et~al.}(2018){Van Eylen}, {Agentoft}, {Lundkvist},
  {Kjeldsen}, {Owen}, {Fulton}, {Petigura}, \& {Snellen}}]{eylen18}
{Van Eylen}, V., {Agentoft}, C., {Lundkvist}, M.~S., {et~al.} 2018, \mnras,
  479, 4786

\bibitem[{{Weiss} \& {Marcy}(2014)}]{weiss14}
{Weiss}, L.~M., \& {Marcy}, G.~W. 2014, \apjl, 783, L6

\bibitem[{{West} {et~al.}(2008){West}, {Hawley}, {Bochanski}, {Covey}, {Reid},
  {Dhital}, {Hilton}, \& {Masuda}}]{west08}
{West}, A.~A., {Hawley}, S.~L., {Bochanski}, J.~J., {et~al.} 2008, \aj, 135,
  785

\bibitem[{{Wu}(2019)}]{wu19}
{Wu}, Y. 2019, \apj, 874, 91

\bibitem[{{Zeng} {et~al.}(2017){Zeng}, {Jacobsen}, \& {Sasselov}}]{zeng17}
{Zeng}, L., {Jacobsen}, S.~B., \& {Sasselov}, D.~D. 2017, Research Notes of the
  American Astronomical Society, 1, 32

\end{thebibliography}
\bibliographystyle{aasjournal}

\end{document}